\documentclass[superscriptaddress,aps,prx,reprint,twocolumn,amsmath,amssymb]{revtex4-2}
\usepackage{graphicx}
\usepackage{hyperref}
\usepackage{amssymb}
\usepackage{slashed}
\usepackage{dcolumn}
\usepackage{amsmath}
\usepackage{bm}
\usepackage{colordvi}
\usepackage{algorithm}
\usepackage{algpseudocode}
\usepackage{multirow}
\usepackage{titlesec}
\usepackage{newtxtext,newtxmath}
\usepackage{dsfont}
\usepackage{xcolor}
\usepackage{mathbbol}

\usepackage{hyperref}
\hypersetup{
    colorlinks=true,
    linkcolor=blue,
    citecolor=blue,     
    urlcolor=blue,
}

\allowdisplaybreaks

\usepackage{mathrsfs}
\makeatletter

\newcommand{\Rmnum}[1]{\expandafter\@slowromancap\romannumeral #1@}
\makeatother

\begin{document}
	\preprint{APS/123-QED}
\title{Preformed excitons in the excitonic density-wave state of monolayer 1$T'$-MoTe$_2$}
\author{Fei Yang}
\thanks{These authors contributed equally to this work.}
\affiliation{Department of Physics, The Hong Kong University of Science and Technology, Hong Kong SAR, China}

\author{Keyu~Wei}
\thanks{These authors contributed equally to this work.}
\affiliation{National Laboratory of Solid State Microstructures and Department of Physics, Nanjing University, Nanjing 210093, China}

\author{Dekun Yang}
\affiliation{Department of Physics, The Pennsylvania State University, University Park, PA 16802, USA}

\author{Kenji~Watanabe}
\affiliation{Research Center for Electronic and Optical Materials, National Institute for Materials Science, 1-1 Namiki, Tsukuba 305-0044, Japan}

\author{Takashi~Taniguchi}
\affiliation{Research Center for Materials Nanoarchitectonics, National Institute for Materials Science, 1-1 Namiki, Tsukuba 305-0044, Japan}

\author{Yong~Zhang}
\affiliation{Jiangsu Provincial Key Laboratory of Advanced Photonic and Electronic Materials, School of Electronic Science and Engineering, Nanjing University, Nanjing 210093, China}

\author{Xuefeng~Wang}
\affiliation{Jiangsu Provincial Key Laboratory of Advanced Photonic and Electronic Materials, School of Electronic Science and Engineering, Nanjing University, Nanjing 210093, China}
\affiliation{Collaborative Innovation Center of Advanced Microstructures, Nanjing University, Nanjing 210093, China}

\author{Fucong~Fei}
\affiliation{National Laboratory of Solid State Microstructures and Department of Physics, Nanjing University, Nanjing 210093, China}

\author{Fengqi~Song}
\affiliation{National Laboratory of Solid State Microstructures and Department of Physics, Nanjing University, Nanjing 210093, China}
\affiliation{Collaborative Innovation Center of Advanced Microstructures, Nanjing University, Nanjing 210093, China}
\affiliation{Jiangsu Physical Science Research Center, Nanjing 210093, China}

\author{Binghai Yan}
\email{binghai.yan@psu.edu}
\affiliation{Department of Physics, The Pennsylvania State University, University Park, PA 16802, USA}
\affiliation{Center for Theory of Emergent Quantum Matter, The Pennsylvania State University, University Park, PA 16802, USA}

\author{Xiaoxiang~Xi}
\email{xxi@nju.edu.cn}
\affiliation{National Laboratory of Solid State Microstructures and Department of Physics, Nanjing University, Nanjing 210093, China}
\affiliation{Collaborative Innovation Center of Advanced Microstructures, Nanjing University, Nanjing 210093, China}
\affiliation{Jiangsu Physical Science Research Center, Nanjing 210093, China}

\author{Junwei Liu}
\email{liuj@ust.hk}
\affiliation{Department of Physics, The Hong Kong University of Science and Technology, Hong Kong SAR, China}

\begin{abstract}
Phase-incoherent preformed excitons have long been anticipated in excitonic insulators, yet their characteristic signatures and underlying physics remain elusive, as the electrical neutrality of excitons hinders direct probes of condensate phase coherence. Here we exploit the emergent excitonic density wave driven by phase-coherent pairings in systems hosting momentum-separated electron and hole Fermi pockets, such as monolayer 1$T'$-MoTe$_2$, as an experimentally accessible probe of excitonic phase coherence. We observe in 1$T'$-MoTe$_2$ that the density-wave order and its associated emergent collective excitations, as revealed by Raman spectroscopy, vanish above $T_c\approx80~$K, whereas the excitonic-insulator state, characterized by a single-particle gap in transport and tunneling measurements, persists up to $T_g\approx100~$K, revealing an intermediate pseudogap-like  regime of preformed excitons without long-range phase coherence. Theoretically, we attribute this intriguing separation to two distinct processes governing the loss of excitonic order by its own emergent excitations: boson-driven phase decoherence and fermion-driven pair breaking. Specifically, we identify an emergent low-lying collective excitation as a lattice-pinned phason (phase mode), whose heating-induced gap softening enhances phase fluctuations and drives condensate phase decoherence above $T_c$, while thermal excitations of fermionic quasiparticles gradually suppress and ultimately destroy pairing amplitude at the higher $T_g$. The  self-consistent microscopic calculation quantitatively captures our key observations, establishing an excitonic realization of preformed-pairing physics.

\end{abstract}

\maketitle

{\sl Introduction.---}Excitonic insulators (EIs)~\cite{PhysRev.158.462,PhysRevLett.19.439,RevModPhys.40.755} constitute a paradigmatic manifestation of interaction-driven quantum order, in which spontaneous electron-hole pairing gives rise to an electronic condensate and opens an interaction-induced gap in the fermionic  quasiparticle spectrum. The associated spontaneous symmetry breaking gives rise to characteristic collective excitations, most notably phase (Nambu-Goldstone~\cite{nambu1960quasi,goldstone1961field,goldstone1962broken,nambu2009nobel,PhysRevX.14.021047,Bretscher2021}) and amplitude (Higgs~\cite{Pekker2015,PhysRevB.101.195118,Werdehausen2018}) modes, which distinguish excitonic condensates from conventional band insulators. Within the mean-field framework, excitonic insulator in systems with substantial conduction-valence band overlap has been viewed as charge-neutral analogues of BCS superconductors~\cite{PhysRev.108.1175,schrieffer1964theory,tinkham2004introduction,abrikosov2012methods,mahan2013many}, with electron-hole pairs replacing Cooper pairs as the elementary constituents of the condensate, exhibiting interaction-driven pair formation and macroscopic quantum coherence.

Many experimentally investigated EI systems~\cite{Sacepe2011,x6k9-wgk9,Rhodes2021,px77-3gg1,PhysRevLett.134.066602,Bi2021,Jia2022,Sun2022,PhysRevLett.125.046801,Varsano2020}, however, are realized in low-dimensional materials. 
A central feature of low-dimensional correlated quantum matter is the possible separation between microscopic  pair formation and the development of global phase coherence~\cite{Emery1995}. For example, in conventional bulk superconductors, Cooper pairing and global phase coherence emerge at nearly the same temperature. In contrast, in two-dimensional (2D) superconductors, global phase coherence can be lost above the superconducting (zero-resistance) transition temperature $T_c$, while microscopic Cooper pairs persist up to a higher temperature $T_g$~\cite{PhysRevB.85.014508,PhysRevLett.106.047001,Charpentier2025,b6vp-zt8z,xy6z-hxcv}, giving rise to a pseudogap regime of phase-incoherent preformed Cooper pairs. This regime in experiments is characterized by a persistent single-particle excitation gap detectable by scanning tunneling microscopy (STM) and angle-resolved photoemission spectroscopy (ARPES), despite the absence of a finite superfluid stiffness (superconductivity) above $T_c$~\cite{Charpentier2025}. 

These phenomena naturally raise the intriguing question of whether an analogous separation between exciton formation and phase coherence can occur in EI  condensates, particularly on the BCS side of the excitonic instability~\cite{Que2026,PhysRevB.85.121102}, where excitonic pairing closely resembles Cooper pairing in superconductors.  However, the electrical neutrality of excitons makes their condensation difficult to probe directly. Accordingly, how phase-incoherent preformed excitons manifest themselves experimentally~\cite{rbvy-4rwd}, and what microscopic mechanisms drive the loss or development of phase coherence of EI pairings and govern the associated phase transitions, remain fundamental open questions. To this end, combining density functional theory (DFT) calculations and self-consistent microscopic theory with spectroscopic measurements, we investigate excitonic ordering in monolayer 1$T'$-MoTe$_2$ and propose a concrete experimental route to probing excitonic phase coherence. We reveal that the formation of electron-hole pairs and the development of excitonic phase coherence occur at distinct temperatures, giving rise to a previously unexplored intermediate regime of phase-incoherent preformed excitons showing pseudogap-like behavior, offering new insights into the EI physics. 

\begin{figure}[t]
\centering
\includegraphics[width=\linewidth]{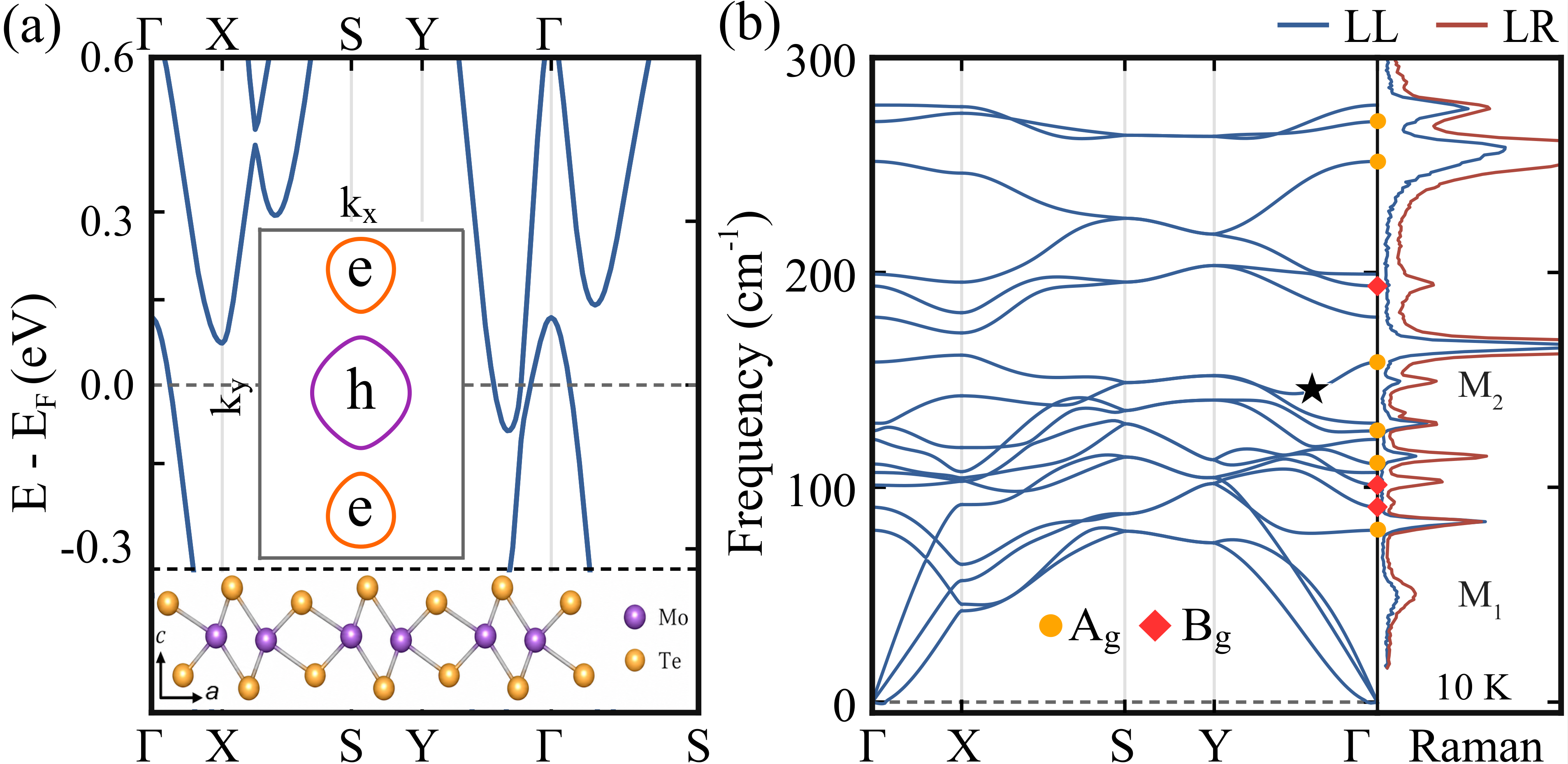}
\caption{(a) DFT-calculated electronic band structure of monolayer 1$T'$-MoTe$_2$, featuring a valence-band pocket centered at $\Gamma$ and two conduction-band pockets at $\pm\mathbf Q_e$, with $\mathbf Q_e=(0,\pi/3b)$. Inset: Electron and hole Fermi-surface pockets.
(b) DFT-calculated phonon band structure compared with polarization-resolved Raman spectra measured at 10~K in the LL and LR configurations, revealing all Raman-active phonon modes at $\Gamma$, along with two additional modes, denoted $M_1$ and $M_2$. Red and orange symbols at $\Gamma$ point indicate the Raman-active $B_g$ and $A_g$ phonon modes, respectively, while the black star marks a  $5b$-folded phonon mode (see Sec.~SIV for details).}
\label{Fig1}
\end{figure}

{\sl Results.---}Monolayer 1$T'$-MoTe$_2$ poesseses a substantial overlap between the conduction and valence bands~\cite{PhysRevB.104.035156}, providing a promising platform for exploring EI  physics. Its low-energy electronic structure consists of a hole pocket centered at the $\Gamma$ point and two symmetry-related electron pockets located at $\pm\mathbf{Q}_e=(0,\pm\pi/3b)$ [Fig.~\ref{Fig1}(a)]. The momentum separation between the electron and hole pockets favors finite-momentum electron-hole pairing, providing a natural mechanism for the formation of an electronically driven excitonic charge-density wave (CDW).  Consistent with this picture, recent low-temperature STM measurements~\cite{x6k9-wgk9,px77-3gg1} revealed a unidirectional density modulation with a periodicity of approximately $5b$, corresponding to an ordering wave vector ${\bf Q}_{\rm CDW}=(0,2\pi/5b)$. This wave vector connects the electron and hole pockets in momentum space~\cite{x6k9-wgk9} and can therefore be written as ${\bf Q}_{\rm CDW}=\mathbf{Q}_e+\mathbf{q}$, where $\mathbf{q}\simeq(k_{F,v}-k_{F,c})\hat{\mathbf y}$ compensates for the mismatch between the electron and hole Fermi surfaces, thereby maximizing the available phase space for excitonic pairing and minimizing the free energy~\cite{yang2026excitonic}.

\begin{figure}[t]
\centering
\includegraphics[width=\linewidth]{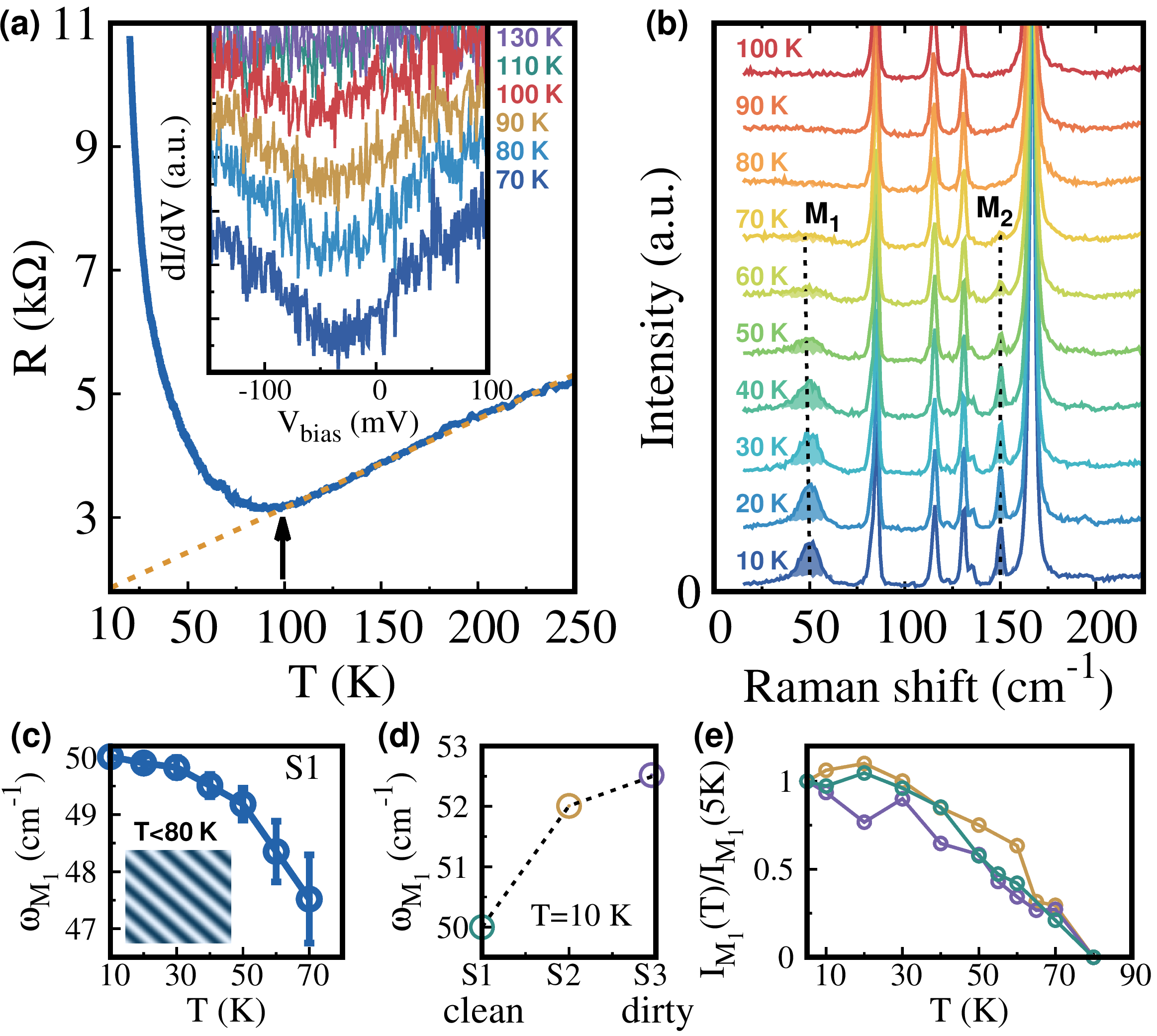}
\caption{(a) Temperature dependence of the electrical resistance. The dashed line indicates a linear fit to the metallic regime, while the arrow marks the onset of deviation from this behavior. Inset: Temperature-dependent differential tunneling conductance after subtraction of the normal-state spectrum at 140~K, $(\mathrm{d}I/\mathrm{d}V)_T-(\mathrm{d}I/\mathrm{d}V)_{140\,\mathrm{K}}$. 
(b) Raman intensity in the $c(bb)\bar{c}$ geometry, clearly showing two additional modes, M$_1$ and M$_2$, that emerge below $T_c=80~$K.
(c) Temperature dependence of the M$_1$ mode frequency. For schematic illustration, we include representative low-temperature STM images showing the $5b$ density modulation, which becomes undetectable above $T\approx80~$K. (d) Raman spectra of the M$_1$ mode in three samples (S1–S3) with progressively increasing disorder induced by successive heating.
(e) Temperature dependence of the M$_1$ Raman intensity, showing that the measured signal disappears near $T_c\approx80$ K in all three samples.}
\label{Fig2}
\end{figure}

To investigate the formation of the EI phase, we experimentally examine the temperature dependence of electrical transport, tunneling spectra, and Raman spectroscopy (see Sec.~SVI for experimental details).  Upon cooling from high temperatures, the electrical resistance [Fig.~\ref{Fig2}(a)] initially exhibits a linear temperature dependence, characteristic of metallic behavior~\cite{Ashcroft1976}, but begins to deviate toward insulating behavior below $T_g\approx100~$K. Consistently, temperature-dependent macroscopic tunneling spectroscopy ($dI/dV$) [inset of Fig.~\ref{Fig2}(a)] reveals a very weak gap-like suppression of low-energy spectral weight that emerges near $T_g\approx100~$K and becomes progressively more pronounced upon cooling, suggesting the gradual opening of a single-particle gap in the fermionic quasiparticle spectrum. Strikingly, as found in Ref.~\cite{Qiu2026}, the characteristic $5b$ density modulation reported in Refs.~\cite{x6k9-wgk9,px77-3gg1} could only be resolved below $\sim80~$K, despite the presence of a well-developed single-particle gap at this temperature [inset of Fig.~\ref{Fig2}(a)]. While the STM may be sensitivity-limited, Raman spectroscopy reveals two additional collective modes (denoted $M_1$ and $M_2$) that are absent from the phonon spectrum predicted by DFT calculations [Fig.~\ref{Fig1}(b)], suggesting their emergence from macroscopic density-wave order. Notably, these modes emerge only below $T_c\approx80~$K [Fig.~\ref{Fig2}(b)], coinciding with the temperature below which the density-wave modulation becomes experimentally resolvable, thereby indicating that the long-range density-wave order sets in at a temperature (80~K) distinctly lower than that  of EI  formation (100~K).

Interestingly, the frequency of $M_2$ remains nearly unchanged with increasing temperature or disorder (see Fig.~SV), suggesting a zone-folded phonon origin as commonly observed in CDW systems~\cite{PhysRevLett.122.127001,Liu2022}, whereas the $M_1$ mode exhibits thermal softening [Fig.~\ref{Fig2}(c)] and systematic disorder-induced hardening [Fig.~\ref{Fig2}(d)], characteristic of a collective excitation of the electronic condensate. Specifically, spontaneous breaking of translational symmetry by a finite-momentum excitonic condensate gives rise to a gapless Nambu-Goldstone phason in the ideal limit~\cite{nambu1960quasi,goldstone1961field,goldstone1962broken,nambu2009nobel,PhysRevX.14.021047,Bretscher2021}, whereas lattice commensurability and disorder pinning in realistic systems can open a gap in its excitation spectrum~\cite{PhysRevB.24.2751,gruner1988dynamics,gruner1985charge,maki1986thermal,PhysRevB.17.535,gpbp-qhp9}, with stronger pinning leading to a higher mode energy. By contrast, the amplitudon describes fluctuations in the order-parameter magnitude~\cite{Pekker2015,PhysRevB.101.195118,Werdehausen2018} and, in electronically driven condensates, typically has an excitation energy of the order of $2|\Delta|$~\cite{Pekker2015} (estimated to be approximately $300~\mathrm{cm}^{-1}$ at $T\sim0$) and can be strongly damped by the  ungapped states~\cite{Shimano2020higgs,Chu2020}. The observed behaviors of $M_1$, including its low-lying excitation energy, thermal softening, and disorder-induced hardening, thus favor its identification as a pinned phason of the EI condensate, whose excitation gap allows long-range phase coherence to persist in 2D, evading the constraint imposed by  Mermin-Wagner theorem.

Given the substantial band overlap in monolayer 1$T'$-MoTe$_2$ which places the system on the BCS side of excitonic condensation~\cite{Que2026,PhysRevB.85.121102}, the intermediate regime between $T_c$ and $T_g$, where the EI  develops without long-range coherent density-wave order, suggests a pseudogap state of phase-incoherent preformed excitons, analogous to preformed Cooper pairs in 2D superconductors~\cite{Sacepe2011,Charpentier2025,PhysRevB.85.014508,PhysRevLett.106.047001,Charpentier2025,b6vp-zt8z,xy6z-hxcv}. Nevertheless, unlike superconductivity, which spontaneously breaks the global $U(1)$ symmetry associated with charge conservation, the excitonic density-wave order possesses a phase degree of freedom that is associated with translational symmetry breaking, corresponding to a phonon-like sliding mode susceptible to pinning, thereby distinguishing its phase dynamics from the BKT picture of vortex-antivortex unbinding in 2D superconductors~\cite{PhysRevB.85.014508,PhysRevLett.106.047001,Charpentier2025,b6vp-zt8z,xy6z-hxcv}. In particular, the characteristic phason energy of $\sim50$ cm$^{-1}$ (corresponding to $\sim72~$K) is comparable to $k_BT_c$. This indicates that the loss of long-range density-wave coherence may be governed by thermal phase fluctuations, arising from thermal excitation of the Nambu-Goldstone phase mode (phason), eventually resulting in a phase-disordered EI state. 

To elucidate the key physics, we develop a self-consistent microscopic theory of the EI formation and phase fluctuations (see Sec.~SI). We begin with the DFT-inspired Hamiltonian
\begin{align}
&H=\sum_{\mathbf k,s}\xi_{v,\mathbf k}
v_{\mathbf k s}^{\dagger}v_{\mathbf k s}
+\sum_{\mathbf k,s,\tau=\pm}
\xi_{c,\mathbf k}
c_{\mathbf k+\tau\mathbf Q_e,s}^{\dagger}
c_{\mathbf k+\tau\mathbf Q_e,s}\nonumber\\
&-\!\sum_{\tau}\!\sum_{\mathbf k,\mathbf k',\mathbf q,s,s'}\!
V^{ss'}_{\mathbf k-\mathbf k'}
c_{\mathbf k+\tau(\mathbf Q_e+\mathbf q),s}^{\dagger}
v_{\mathbf k,s'}
v_{\mathbf k',s'}^{\dagger}
c_{\mathbf k'+\tau(\mathbf Q_e+\mathbf q),s}.\label{Ham}
\end{align}
Here, $\xi_{c,\mathbf k}$ and $\xi_{v,\mathbf k}$ denote the conduction- and valence-band dispersions measured relative to their respective pocket centers, and $V^{ss'}_{\mathbf k-\mathbf k'}$ describes the effective electron-hole attraction.  In the large-band-overlap regime, corresponding to the BCS side of the excitonic instability, we focus on the leading $s$-wave pairing channel and approximate the interaction as $V^{ss'}_{\mathbf{k}-\mathbf{k}'}\simeq g\delta_{ss'}$ following conventional  treatments~\cite{PhysRev.158.462,PhysRevLett.19.439,RevModPhys.40.755,Bi2021,PhysRevB.71.224502}. Time-reversal symmetry relates the two valley-selective excitonic pairing channels, such that valence-band states at $k_y>0$ ($k_y<0$) pair with conduction-band states near $+\mathbf Q_e$ ($-\mathbf Q_e$), respectively, preserving 
the overall time-reversal symmetry of the system. The corresponding EI order parameter reads
\begin{equation}
\Delta_{\mathbf q}
=g\sum_{\mathbf k}\langle
c_{\mathbf k+(\mathbf Q_e+\mathbf q),s}^{\dagger}
v_{\mathbf k,s}\rangle. 
\label{eq:Delta_eta}
\end{equation}
Hereafter, $\mathbf{k}$ is restricted to $k_y>0$, with the $k_y<0$ contribution obtained by time reversal. The additional pairing momentum $\mathbf q$ is introduced to compensate for the electron-hole Fermi-surface mismatch and maximize the available phase space for pairing, similar to a FFLO state~\cite{fulde1964superconductivity,larkin1965zh,PhysRevA.89.013607,PhysRevLett.89.227002,PhysRevB.76.014522,PhysRevB.75.064511,PhysRevLett.114.110401}. 
This excitonic condensate naturally yields a density-wave modulation,
\begin{align}
&\delta\rho(\mathbf R)
=2\sum_{\mathbf k,s}\langle
c_{\mathbf k+{\bf Q}_{\rm CDW},s}^{\dagger}
v_{\mathbf k,s}\rangle
e^{i{\bf Q}_{\rm CDW}\cdot\mathbf R}\!+\!\mathrm{c.c.}\nonumber\\
&=\!\frac{2\Delta_{\bf q}}{g}
e^{i{\bf Q}_{\rm CDW}\cdot\mathbf R}
\!+\!\mathrm{c.c.}\!=\!
\frac{4|\Delta_{\bf q}|}{g}
\cos\left(
{\bf Q}_{\rm CDW}\cdot\mathbf R
\!+\!\theta_0\!+\!\delta\theta\right),\label{de}
\end{align}
with ${\bf Q}_{\rm CDW}=\mathbf Q_e+\mathbf q$. Here, we take $\Delta_{\mathbf q}=|\Delta_{\mathbf q}|e^{i(\theta_0+\delta\theta)}$, where $|\Delta_{\mathbf q}|$ denotes the pairing amplitude, $\theta_0$ is the equilibrium phase, and $\delta\theta$ describes collective phase fluctuations. Within the self-consistent harmonic approximation (SCHA) with $\langle\delta\theta\rangle=0$ and $\langle(\delta\theta)^2\rangle\neq0$, averaging over   fluctuations using a cumulant expansion yields the observable modulation (see Sec.~SI~B),
\begin{equation}
\delta{\bar\rho}({\bf R})
=
\frac{4|\Delta_{\bf q}|}{g}
\exp\big[
-\frac12\left\langle(\delta\theta)^2\right\rangle\big]\cos\left(
{\bf Q}_{\rm CDW}\cdot\mathbf R\!+\!\theta_0
\right).
\label{eq:rho_eff}
\end{equation}
The exponential factor describes the suppression (phase decoherence) of the density-wave modulation by phase fluctuations, allowing long-range density-wave coherence to vanish even when the excitonic pairing amplitude remains finite.

Applying the mean-field framework to Hamiltonian~(\ref{Ham}) with Eq.~(\ref{eq:Delta_eta})  yields the quasiparticle energies (see Sec.~SI~A, including the reconstructed Fermi surfaces and EI correlations),
\begin{equation}
E_{\mathbf k}^{\pm}
=\frac{\xi_{c,\mathbf k+\mathbf q}+\xi_{v,\mathbf k}}{2}\pm 
\sqrt{\big(
\frac{\xi_{c,\mathbf k+\mathbf q}-\xi_{v,\mathbf k}}{2})^2
+
|\Delta_{\mathbf q}|^2
},\label{qe}
\end{equation}
and the self-consistent equation of pairing amplitude $|\Delta_{\bf q}|$,
\begin{equation}
\frac{1}{g}
=-
\sum_{\mathbf k}
\frac{
f(E_{\mathbf k}^{+})-f(E_{\mathbf k}^{-})
}{E_{\mathbf k}^{+}-E_{\mathbf k}^{-}},
\label{eq:gap}
\end{equation}
where $f(E)$ is the Fermi  distribution. To incorporate thermal phase fluctuations, we derive the effective action governing the phase dynamics within the imaginary-time path-integral formalism (Sec.~SI~C1), yielding the phason dispersion:
\begin{equation}
\omega_{\mathbf p}
=
\sqrt{
m_{\theta}^2+{f^c_{ij}p_ip_j}/{D}
}.
\label{eq:phason_dispersion}
\end{equation}
where $D$ and $f^c_{ij}=\sum_{\mathbf k}\rho_{ij,\mathbf k}(|\Delta_{\bf q}|)$ denote the temporal and spatial phase stiffness, respectively, self-consistently determined by  fermionic response (see Sec.~SI~C); $m_\theta$ characterizes the pinning strength of the phason induced by lattice commensurability, giving rise to a finite phason gap, and self-consistently accounting for the suppression of the pinning potential by thermal phase fluctuations yields (see Sec.~SI C2)
\begin{equation}
m_{\theta}^2(T)
=
\frac{4l^2\lambda}{D}
\exp\big[
-\frac{{l}^2}{2}\left\langle(\delta\theta)^2\right\rangle\big],
\label{eq:phasonmass}
\end{equation}
where $\lambda$ is the pinning strength and $l=5$ (commensurability). Within  quantum statistics, the thermal phase fluctuations follows a standard bosonic thermal excitation (see Sec.~SI~C3), 
\begin{equation}
\Big\langle
\Big(\frac{\delta\theta}{2}\Big)^2
\Big\rangle_{\rm th}
=
\int\frac{d^2p}{(2\pi)^2}
\frac{2n_B(\omega_{\mathbf p})}
{2D\omega_{\mathbf p}},
\label{eq:phase_variance}
\end{equation}
where $n_B(\omega)$ is the Bose-Einstein distribution. Equations~(\ref{de})-(\ref{eq:phase_variance}) are solved self-consistently: the equilibrium pairing amplitude $|\Delta(T)|$ and momentum $\mathbf q_0$ are obtained by maximizing the self-consistent solution $|\Delta_{\mathbf q}|$ [Eq.~(\ref{eq:gap})] over $\mathbf q$,  equivalent to minimizing  free energy with respect to $|\Delta_{\mathbf q}|$ and $\mathbf q$~\cite{yang2026excitonic}; the resulting $|\Delta(T)|$ determines the phase stiffness $f^c_{ij}(T)$, while phase fluctuations renormalize the pinning gap in a self-consistent manner and suppress the coherent density-wave order.

\begin{figure}[t]
\centering
\includegraphics[width=\linewidth]{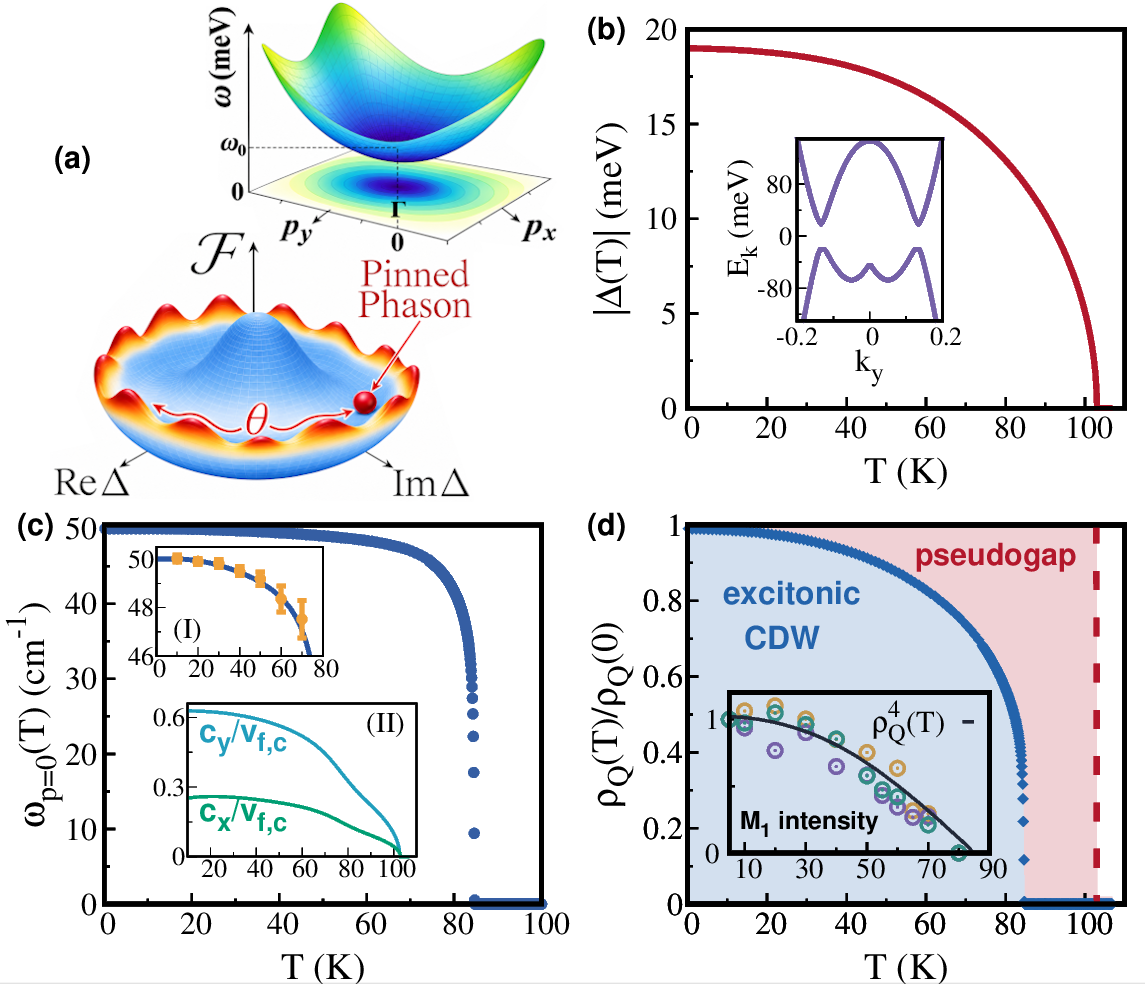}
\caption{(a) Schematic illustration of the free-energy landscape of the complex order parameter $\Delta$, and the energy dispersion $\omega_{\bf p}$ of a pinned phason with a finite excitation gap. (b) Pairing amplitude $|\Delta(T)|$, which vanishes at $T_g=104~$K. Inset: Fermionic quasiparticle spectrum $E_{\bf k}^{\pm}$  [Eq.~(\ref{qe})] along the $k_y$ direction at 10~K, illustrating the opening of the single-particle gap. (c) Phason gap $\omega_{{\bf p}=0}(T)$. Inset (I): Comparison of the calculated phason gap with Raman measurements. Inset (II): Phason velocities $c_{i=x,y}(T)=\sqrt{f^c_{ii}(T)/D}$. (d) The calculated density-wave order $\rho_Q(T)=2|\Delta|\exp[-\langle(\delta\theta)^2\rangle/2]$ and the phase diagram. The dashed line marks $T_g$, where $|\Delta(T)|$ vanishes, while $\rho_Q(T)$ vanishes at $T_c=84~$K.  Inset: The normalized  $M_1$ Raman intensity $I_{M_1}(T)/I_{M_1}(5~\text{K})$ measured for three samples (S1, S2, and S3) compared with a phenomenological scaling $I_{M_1}\propto\rho_{\mathrm{Q}}^4(T)$~\cite{EIdiscussion}.}
\label{Fig3}
\end{figure}

For the numerical calculations, we approximate the conduction and valence bands by isotropic parabolic dispersions, with the band overlap, chemical potential, and effective masses determined from DFT calculations. Based on experimental measurements, we estimate the zero-temperature excitonic gap to be $2|\Delta(0)|\approx 38$ meV, which determines the effective electron-hole interaction strength, and take $\sqrt{4l^2{\lambda}/D}=50~\mathrm{cm}^{-1}$ as the pinning strength. The resulting self-consistent calculation results are presented in Fig.~\ref{Fig3}.  The calculations yield a finite pairing momentum $\mathbf q_0=(k_{F,v}-k_{F,c})\hat{\mathbf y}=(0,0.05)~\mathrm{\AA}^{-1}$, which compensates for the Fermi-surface mismatch between the electron and hole pockets. The resulting density-wave ordering vector remains nearly constant across the entire temperature range of the phase, with ${\bf Q}_{\rm CDW}\approx(0,0.35)~\mathrm{\AA}^{-1}$, close to the experimental value $\mathbf Q_{\rm CDW}^{\rm exp}\approx(0,2\pi/5b)=(0,0.36)~\mathrm{\AA}^{-1}$.

The obtained low-temperature quasiparticle spectrum [inset of Fig.~\ref{Fig3}(b)] exhibits the characteristic gap opening expected for an EI formation, consistent with the insulating behavior observed in electrical resistance measurements. Self-consistent numerical solutions of the gap equation [Eq.~(\ref{eq:gap})], using the material parameters described above, show that the single-particle gap $|\Delta(T)|$ gradually decreases with increasing temperature and vanishes at $T_g=104~\mathrm{K}$ [Fig.~\ref{Fig3}(b)]. This temperature evolution originates from the thermal excitation of fermionic quasiparticles, which progressively suppresses electron-hole pairing and ultimately drives the pairing amplitude $|\Delta|$ to zero, restoring the metallic state above $T_g$ as observed in Fig.~\ref{Fig2}(a).

Meanwhile, the phason dispersion $\omega_{\mathbf p}$ exhibits clear  thermal softening behavior: as temperature increases, its excitation gap $\omega_{p=0}(T)$ [Fig.~\ref{Fig3}(c)] is self-consistently renormalized downward by thermal phase fluctuations due to enhanced bosonic excitations, whereas its velocity  [inset (II) of Fig.~\ref{Fig3}(c)] decreases as the phase stiffness $f_{ij}^c(T)$ is reduced due to the suppressed $|\Delta(T)|$. The calculated temperature dependence of the phason gap quantitatively agrees with the experimental measurements of M$_1$ [inset (I) of Fig.~\ref{Fig3}(c)], supporting its identification as a pinned phason of the EI condensate. Notably, despite assuming isotropic electronic band dispersions, our results reveal an anisotropic phase stiffness, with $f^c_{yy}>f^c_{xx}$, indicating an emergent excitonic nematicity. This anisotropy originates from the pairing momentum $\mathbf q_0=(k_{F,v}-k_{F,c})\hat{\mathbf y}$, which compensates for the Fermi-surface mismatch between the electron and hole pockets. Accordingly, electron-hole pairing correlations are predominantly concentrated along the $y$ direction, resulting in a larger phase stiffness along $y$ and, consequently, a higher phason velocity along $y$ than along $x$.

The critical regime is reached when $k_{\mathrm B}T$ becomes comparable to the phason gap $\omega_{\mathbf p=0}$. The resulting increase in the thermal population of low-energy phason modes, $n_{\mathrm B}(\omega_{\mathbf p\sim0})$, enhances phase fluctuations, which further renormalize the pinning gap downward. This positive feedback drives a rapid, self-amplifying increase in phase fluctuations, causing the phason excitation gap $m_{\theta}(T)$ to collapse and the coherent density-wave order to vanish abruptly at approximately $84~\mathrm{K}$, well below the single-particle gap-closing temperature $T_g=104~\mathrm{K}$. Notably, above $80~\mathrm{K}$, the system enters a phase-disordered regime in which long-range phase coherence is lost and the density-wave order vanishes, while the single-particle excitation gap remains finite, consistent with its persistence in tunneling spectroscopy and the insulating behavior observed experimentally. This regime thus exhibits preformed excitonic pairing and pseudogap behavior. Importantly, even in this regime, the phason becomes effectively gapless while retaining a finite velocity  [inset (II) of Fig.~\ref{Fig3}(c)], which vanishes only as the excitonic gap $|\Delta(T)|$ closes near $T_g$. The persistence of a dispersive phase mode reflects the collective phase dynamics associated with excitonic pairing, whose existence is guaranteed by the Nambu-Goldstone theorem in the presence of spontaneously broken continuous symmetry~\cite{nambu1960quasi,goldstone1961field,goldstone1962broken,nambu2009nobel}.

{\sl Discussion.---}Preformed excitons are commonly discussed in the BEC regime~\cite{Que2026,PhysRevB.85.121102,Kaneko2025}, where tightly bound electron-hole pairs form above the condensation temperature. Our results reveal a long-overlooked possibility of preformed excitons on the BCS side, where a large band overlap and substantial Fermi surfaces persist. In this regime, low dimensionality amplifies thermal fluctuations of the Goldstone phase mode, allowing long-range phase coherence to disappear while the excitonic pairing amplitude remains finite above the density-wave ordering temperature.  This separation between pairing and phase coherence reveals that preformed excitons do not necessarily require tightly bound electron-hole pairs characteristic of the BEC regime, but can also emerge from phase-disordered excitonic pairing in a BCS-side  electronic structure with substantial band overlap and well-defined Fermi surfaces. 

Typically, probing phase coherence in EIs is inherently challenging because excitons are electrically neutral. In monolayer $1T'$-MoTe$2$, however, phase-coherent intervalley electron-hole pairing gives rise to an excitonic density wave, providing accessible signatures of excitonic phase coherence. The distinct temperature evolutions of the excitonic pairing amplitude and long-range phase coherence enable us to explore an intermediate pseudogap regime of phase-incoherent preformed excitons, and reveal how excitonic order is progressively destroyed by its own emergent excitations (bosonic phasons and fermionic quasiparticles) via distinct microscopic mechanisms: boson-driven phase decoherence and fermion-driven pair breaking. 
 These findings reveal a rich landscape of ordering and disordering phenomena in the interacting many-body systems.

{\sl Acknowledgments.---}J.L. acknowledge support from the Hong Kong Research Grants Council (CRS\_HKUST603/25, C6046-24G, 16304523, 16311125, 16311226, and AoE/P-604/25R). B.Y. acknowledges the financial support by the National Science Foundation (NSF) DMREF Award 2522898 and DMREF Award 2522897.  X.X. acknowledges support from the National Natural Science Foundation of China (Grants No.~U24A6002 and No.~12474170), the Fundamental Research Funds for the Central Universities (Grant No.~0204-14380264), and the Natural Science Foundation of Jiangsu Province (Grant No.~BK20233001). K.W. and T.T. acknowledge support from the CREST (JPMJCR24A5), JST and World Premier International Research Center Initiative (WPI), MEXT, Japan.

\makeatletter
\renewcommand{\bibsection}{%
    \par
    \begingroup
    \baselineskip26\p@
    \bib@device{\hsize}{72\p@}%
    \endgroup
    \nobreak\@nobreaktrue
    \addvspace{19\p@}%
}
\makeatother

\bibliography{ref.bib}

\end{document}


\title{Preformed excitons in the excitonic density-wave state of monolayer 1$T'$-MoTe$_2$ (Supplementary  Materials)}

\author{Fei Yang}
\thanks{These authors contributed equally to this work.}
\affiliation{Department of Physics, The Hong Kong University of Science and Technology, Hong Kong SAR, China}

\author{Keyu~Wei}
\thanks{These authors contributed equally to this work.}
\affiliation{National Laboratory of Solid State Microstructures and Department of Physics, Nanjing University, Nanjing 210093, China}

\author{Dekun Yang}
\affiliation{Department of Physics, The Pennsylvania State University, University Park, PA 16802, USA}

\author{Kenji~Watanabe}
\affiliation{Research Center for Electronic and Optical Materials, National Institute for Materials Science, 1-1 Namiki, Tsukuba 305-0044, Japan}

\author{Takashi~Taniguchi}
\affiliation{Research Center for Materials Nanoarchitectonics, National Institute for Materials Science, 1-1 Namiki, Tsukuba 305-0044, Japan}

\author{Yong~Zhang}
\affiliation{Jiangsu Provincial Key Laboratory of Advanced Photonic and Electronic Materials, School of Electronic Science and Engineering, Nanjing University, Nanjing 210093, China}

\author{Xuefeng~Wang}
\affiliation{Jiangsu Provincial Key Laboratory of Advanced Photonic and Electronic Materials, School of Electronic Science and Engineering, Nanjing University, Nanjing 210093, China}
\affiliation{Collaborative Innovation Center of Advanced Microstructures, Nanjing University, Nanjing 210093, China}

\author{Fucong~Fei}
\affiliation{National Laboratory of Solid State Microstructures and Department of Physics, Nanjing University, Nanjing 210093, China}

\author{Fengqi~Song}
\affiliation{National Laboratory of Solid State Microstructures and Department of Physics, Nanjing University, Nanjing 210093, China}
\affiliation{Collaborative Innovation Center of Advanced Microstructures, Nanjing University, Nanjing 210093, China}
\affiliation{Jiangsu Physical Science Research Center, Nanjing 210093, China}

\author{Binghai Yan}
\email{binghai.yan@psu.edu}
\affiliation{Department of Physics, The Pennsylvania State University, University Park, PA 16802, USA}
\affiliation{Center for Theory of Emergent Quantum Matter, The Pennsylvania State University, University Park, PA 16802, USA}

\author{Xiaoxiang~Xi}
\email{xxi@nju.edu.cn}
\affiliation{National Laboratory of Solid State Microstructures and Department of Physics, Nanjing University, Nanjing 210093, China}
\affiliation{Collaborative Innovation Center of Advanced Microstructures, Nanjing University, Nanjing 210093, China}
\affiliation{Jiangsu Physical Science Research Center, Nanjing 210093, China}

\author{Junwei Liu}
\email{liuj@ust.hk}
\affiliation{Department of Physics, The Hong Kong University of Science and Technology, Hong Kong SAR, China}

\maketitle 

\tableofcontents

\begin{figure}[h]
\centering
\includegraphics[width=\linewidth]{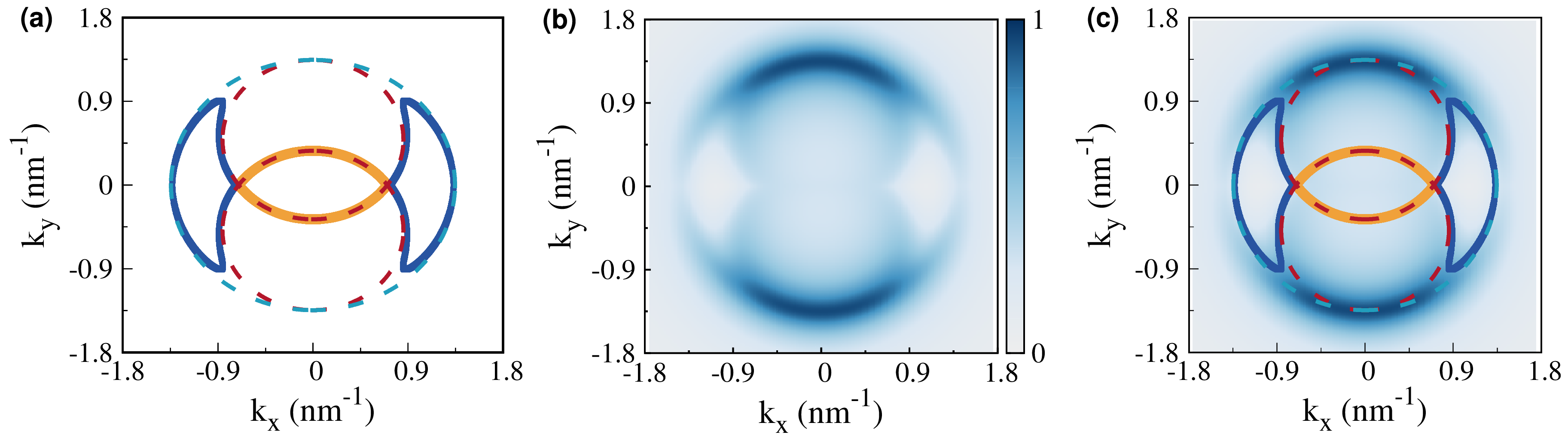}
\caption{(a) Theoretically calculated Fermi surfaces at 10 K. Dashed curves show the folded Fermi surfaces in the absence of exciton-insulator (EI) formation, with red and blue denoting the electron ($\xi_{c,\mathbf{k}\pm\mathbf{q}}=0$) and hole ($\xi_{v,\mathbf{k}}=0$) pockets, respectively. Solid curves show the reconstructed Fermi surfaces in the EI state, where blue denotes the reconstructed quasiparticle bands and yellow indicates the residual portions of the original bands. The overlapping regions of the original electron and hole Fermi surfaces are gapped out by EI formation. (b) Momentum-resolved EI correlations projected onto the valence band, showing that  electron-hole pairing correlations are predominantly concentrated in the regions where the original electron and hole Fermi surfaces overlap. (c) Overlay of (a) and (b), demonstrating that the strongest EI correlations coincide with the regions where the electron and hole Fermi surfaces are matched and subsequently gapped out, as expected in the EI formation.}
\label{FS_LO}
\end{figure}

\section{Self-consistent microscopic theoretical framework of phase-fluctuating exciton insulators}

To elucidate the underlying physics, we develop a self-consistent microscopic theory that treats fermionic quasiparticles and bosonic collective phase modes on equal footing. The resulting simulation  quantitatively reproduces both characteristic temperatures: it captures the loss of long-range density-wave order (phase coherence) at approximately $80~\mathrm{K}$ due to thermally excited bosonic phasons, while the EI pairing persists up to $100~\mathrm{K}$ before being destroyed by fermionic quasiparticle excitations,  revealing a microscopic distinction between boson-driven phase decoherence and fermion-driven pair breaking in 2D EIs. 

\subsection{Mean-field Hamiltonian, gap equation, and fermionic quasiparticle spectrum}
\label{SMham}

The DFT-calculated phonon dispersion of monolayer 1$T'$-MoTe$_2$ exhibits neither imaginary frequencies nor low-lying soft optical modes  [Fig.~1(b) of the main text], indicating the dynamical stability of the lattice. On the electronic side, the substantial overlap between the conduction and valence bands provides favorable conditions for excitonic instability on the BCS side~\cite{Que2026,PhysRevB.85.121102}. In particular, the momentum separation between the electron and hole pockets favors finite-momentum electron-hole pairing, offering a natural mechanism for the formation of an electronically driven excitonic CDW. To develop a self-consistent microscopic theory, we start with a microscopic interacting Hamiltonian based on the DFT-calculated electronic band structure of monolayer 1$T'$-MoTe$_2$, featuring a hole pocket centered at $\Gamma$ and two electron pockets located at $\pm\mathbf Q_e$, 
\begin{equation}
H=\sum_{\mathbf k,s}\xi_{v,\mathbf k}
v_{\mathbf k s}^{\dagger}v_{\mathbf k s}
+\sum_{\mathbf k,s,\tau=\pm}
\xi_{c,\mathbf k}
c_{\mathbf k+\tau\mathbf Q_e,s}^{\dagger}
c_{\mathbf k+\tau\mathbf Q_e,s}\!-\!\sum_{\tau}\!\sum_{\mathbf k,\mathbf k',\mathbf q,s,s'}\!
V^{ss'}_{\mathbf k-\mathbf k'}
c_{\mathbf k+\tau(\mathbf Q_e+\mathbf q),s}^{\dagger}
v_{\mathbf k,s'}
v_{\mathbf k',s'}^{\dagger}
c_{\mathbf k'+\tau(\mathbf Q_e+\mathbf q),s}.
\label{eq:SM_Hamiltonian}
\end{equation}
Here, $\xi_{v,\mathbf k}$ and $\xi_{c,\mathbf k}$ describe the valence- and conduction-band dispersions measured relative to their respective pocket centers, while $V_{\mathbf k-\mathbf k'}^{ss'}$ represents the effective electron-hole attraction. The substantial overlap between the conduction and valence bands places the system in the BCS regime of electron-hole pairing, where electron-hole correlations predominantly develop near the Fermi surfaces. Within the conventional exciton-insulator framework, the dominant $s$-wave pairing channel is considered, allowing the interaction to be approximated by a momentum-independent coupling, $V_{\mathbf k-\mathbf k'}^{ss'}\simeq g\delta_{ss'}$, with $g>0$~\cite{PhysRev.158.462,PhysRevLett.19.439,RevModPhys.40.755,Bi2021,PhysRevB.71.224502}. The presence of two symmetry-related electron valleys gives rise to two distinct electron-hole pairing channels, with valence-band states at $k_y>0$ ($k_y<0$) preferentially pairing with conduction-band states near $+\mathbf Q_e$ ($-\mathbf Q_e$). These two channels are related by time-reversal symmetry, allowing the electronic condensate to preserve the overall time-reversal symmetry of the system. Accordingly, the exciton-insulator order parameter in monolayer 1$T'$-MoTe$_2$ is defined as~\cite{PhysRev.158.462,PhysRevLett.19.439,RevModPhys.40.755,Bi2021,PhysRevB.71.224502} 
\begin{equation}
\Delta_{\mathbf q}
=g\sum_{\mathbf k,s}\langle
c_{\mathbf k+(\mathbf Q_e+\mathbf q),s}^{\dagger}
v_{\mathbf k,s}\rangle,  
\label{eq:SM_Delta}
\end{equation}
Here and throughout the following calculations, the momentum ${\bf k}$ is restricted to $k_y>0$, while the contribution from $k_y<0$ is obtained by time-reversal symmetry. The additional pairing momentum $\mathbf q$ is introduced to accommodate the substantial mismatch between the electron and hole Fermi surfaces. Specifically, the conduction-band Fermi momentum is smaller than that of the valence band in monolayer 1$T'$-MoTe$_2$, $k_{F,c}<k_{F,v}$, such that folding the two pockets by $\mathbf Q_e$ alone cannot provide Fermi-surface overlap. A finite $\mathbf q$ therefore improves their momentum-space matching, enlarging the available phase space for electron-hole pairing and lowering the free energy, analogous to the FFLO  state in superconductors~\cite{fulde1964superconductivity,larkin1965zh,yang2018fulde,Dong_2013,PhysRevA.89.013607,PhysRevLett.89.227002,PhysRevB.76.014522,PhysRevB.75.064511,PhysRevLett.114.110401}.  

Within the mean-field framework~\cite{mahan2013many,abrikosov2012methods}, the interacting Hamiltonian reduces to
\begin{align}
H_{\mathrm{MF}}
=&\sum_{\mathbf k,s}
(c^{\dagger}_{\mathbf k+(\mathbf Q_e+\mathbf q),s}, v^{\dagger}_{\mathbf k,s})
\begin{pmatrix}
\xi_{c,\mathbf k+\mathbf q} & -\Delta_{\mathbf q}\\
-\Delta_{\mathbf q}^{*} & \xi_{v,\mathbf k}
\end{pmatrix}
\begin{pmatrix}
c_{\mathbf k+(\mathbf Q_e+\mathbf q),s}\\
v_{\mathbf k,s}
\end{pmatrix}
+\frac{|\Delta_{\mathbf q}|^2}{g}.
\label{eq:SM_HMF}
\end{align}
Diagonalization of this mean-field Hamiltonian gives the fermionic quasiparticle energies
\begin{equation}
E_{\mathbf k}^{\pm}
=\frac{\xi_{c,\mathbf k+\mathbf q}+\xi_{v,\mathbf k}}{2}\pm 
\sqrt{\big(
\frac{\xi_{c,\mathbf k+\mathbf q}-\xi_{v,\mathbf k}}{2})^2
+
|\Delta_{\mathbf q}|^2
}.
\label{eq:SM_Ek}
\end{equation}
The pairing amplitude $|\Delta_{\mathbf q}|$ is determined self-consistently by the gap equation~\cite{mahan2013many,abrikosov2012methods,fulde1964superconductivity,larkin1965zh,yang2018fulde,Dong_2013,PhysRevA.89.013607,PhysRevLett.89.227002,PhysRevB.76.014522,PhysRevB.75.064511,PhysRevLett.114.110401}
\begin{equation}
\frac{1}{g}
=
\sum_{\mathbf k}
\frac{
f(E_{\mathbf k}^{-})-f(E_{\mathbf k}^{+})
}{
E_{\mathbf k}^{+}-E_{\mathbf k}^{-}
},
\label{eq:SM_gap}
\end{equation}
where $f(E)$ is the Fermi-Dirac distribution function. For each $\mathbf q$, the gap equation determines the corresponding self-consistent pairing amplitude $|\Delta_{\mathbf q}(T)|$. The equilibrium pairing momentum $\mathbf q_0$ and pairing amplitude $\Delta(T)$ are then determined by
\begin{equation}
\begin{aligned}
\mathbf q_0=\underset{\mathbf q}{\operatorname{arg\,max}}\,
|\Delta_{\mathbf q}(T)|,\qquad
|\Delta(T)|=|\Delta_{\mathbf q_0}(T)|.
\end{aligned}
\label{eq:SM_q0}
\end{equation}
Maximizing the self-consistent gap magnitude, as established in Ref.~\cite{yang2026excitonic}, is equivalent to minimizing the mean-field free energy $\Omega(|\Delta_{\bf q}|,{\bf q})$ with respect to both the pairing amplitude $|\Delta_{\mathbf q}|$ and momentum $\mathbf q$, 
\begin{equation}
\partial_{|\Delta_{\bf q}|}\Omega(|\Delta_{\bf q}|,{\bf q})=0,\quad \partial_{\bf q}\Omega(|\Delta_{\bf q}|,{\bf q})=0.
\end{equation}\\

The self-consistent calculations, as detailed in Sec.~\ref{sec_sd}, yield the optimal pairing momentum $\mathbf q_0=(k_{F,v}-k_{F,c})\hat{\mathbf y}$ and hence the total ordering vectors $\mathbf Q_{\rm CDW}=\mathbf Q_e+\mathbf q_0$. Physically, folding the original Fermi-surface pockets shown in the inset of Fig.~2(b) of the main text by $\mathbf Q_{\rm CDW}$ yields the dashed contours in Fig.~\ref{FS_LO}(a), where the resulting Fermi-surface overlap enlarges the available phase space for electron-hole pairing, thereby lowering the free energy. Consequently, a finite pairing amplitude $|\Delta|$ opens a gap in the quasiparticle spectrum $E_{\mathbf k}^{\pm}$ [see the inset of Fig.~3(b) in the main text], gapping out the electronic states in the overlapping regions of the folded electron and hole Fermi surfaces,  as illustrated by the resulting reconstructed quasiparticle Fermi surfaces (solid blue contours determined by $E_{\mathbf k}^{\pm}=0$) in Fig.~\ref{FS_LO}(a). Meanwhile, the conduction-band states outside the valley-selective pairing channels retain their original dispersion, forming residual Fermi surfaces (solid yellow contours). Notably, both the reconstructed quasiparticle and residual conduction-band Fermi surfaces form closed contours. The momentum-resolved electron-hole pairing correlation at the equilibrium pairing momentum is given by
\begin{equation}
F_{\rm EI}(\mathbf k)
=\sum_s\langle
c_{\mathbf k+(\mathbf Q_e+\mathbf q_0),s}^{\dagger}
v_{\mathbf k,s}\rangle
=
\left.
\frac{|\Delta_{\mathbf q}|\left[f(E_{\mathbf k}^{-})-f(E_{\mathbf k}^{+})\right]}
{E_{\mathbf k}^{+}-E_{\mathbf k}^{-}}
\right|_{\mathbf q=\mathbf q_0}.
\label{eq:SM_EI_correlation}
\end{equation}
As shown in Fig.~\ref{FS_LO}(b), $F_{\rm EI}(\mathbf k)$ exhibits a pronounced momentum-space anisotropy, with strong pairing correlations concentrated along the $k_y$ direction and nearly vanishing correlations along the $k_x$ direction.  Figure~\ref{FS_LO}(c) clearly demonstrates that the strongest  electron-hole pairing correlations occur in the regions where the electron and hole Fermi surfaces overlap, precisely coinciding with the electronic states that are gapped out by electron-hole pairing.

\subsection{Excitonic density-wave order}

As an electronic condensate, the excitonic-insulator order parameter takes a generalized complex form~\cite{abrikosov2012methods,nambu1960quasi,nambu2009nobel} 
\begin{align}
\Delta_{\mathbf q}
=
|\Delta_{\mathbf q}|
e^{i(\theta_0+\delta\theta)},
\end{align}
where $|\Delta_{\mathbf q}|$ denotes the electron-hole pairing amplitude determined by the self-consistent gap equation above, $\theta_0$ is the equilibrium phase, and $\delta\theta$ describes the collective phase fluctuations. The intervalley finite-momentum excitonic condensate in Eq.~(\ref{eq:SM_Delta}) then naturally gives rise to a density-wave modulation~\cite{PhysRev.158.462,PhysRevLett.19.439,RevModPhys.40.755,Bi2021,PhysRevB.71.224502},
\begin{align}
\delta\rho(\mathbf R)=2
\sum_{\mathbf k,s}\langle
c_{\mathbf k+\mathbf Q_{\rm CDW},s}^{\dagger}
v_{\mathbf k,s}\rangle
e^{i\mathbf Q_{\rm CDW}\cdot\mathbf R}
+\mathrm{c.c.}=\frac{2\Delta_{\bf q}}{g}
e^{i{\bf Q}_{\rm CDW}\cdot\mathbf R}
\!+\!\mathrm{c.c.}=\frac{4|\Delta_{\mathbf q}|}{g}
\cos\left(
\mathbf Q_{\rm CDW}\cdot\mathbf R
+\theta_0+\delta\theta
\right),
\end{align}
where $\mathbf Q_{\rm CDW}=\mathbf Q_e+\mathbf q$. Within the self-consistent harmonic approximation (SCHA),  the thermal phase fluctuations satisfy $\langle\delta\theta\rangle=0$ and $\langle[\delta\theta]^2\rangle\neq0$, allowing the thermal average to be evaluated using the cumulant expansion~\cite{peskin2018introduction},
\begin{align}
\left\langle\cos(\theta_e+\delta\theta)\right\rangle
=
\mathrm{Re}\left[
e^{i\theta_e}
\left\langle e^{i\delta\theta}\right\rangle
\right]&=
\mathrm{Re}\Big[
e^{i\theta_e}
\sum_{n=0}^{\infty}
\frac{(-1)^n\langle(\delta\theta)^{2n}\rangle}{(2n)!}
\Big]=
\mathrm{Re}\Big[
e^{i\theta_e}
\sum_{n=0}^{\infty}
\frac{(-1)^n(2n-1)!!\langle(\delta\theta)^2\rangle^n}{(2n)!}
\Big]\nonumber\\
&=
\mathrm{Re}\Big[
e^{i\theta_e}
\sum_{n=0}^{\infty}
\frac{(-1)^n\langle(\delta\theta)^2\rangle^n}{2^n n!}
\Big]=
\cos\theta_e
\exp\Big[
-\frac12\langle(\delta\theta)^2\rangle
\Big],
\label{DW}
\end{align}
where $\theta_e=\mathbf Q_{\rm CDW}\cdot\mathbf R+\theta_0$. Accordingly,  averaging over thermal phase fluctuations   yields the observable  modulation,
\begin{equation}
\delta{\bar\rho}({\bf R})=
\frac{4|\Delta_{\mathbf q}|}{g}
\exp\Big[
-\frac{1}{2}
\left\langle(\delta\theta)^2\right\rangle
\Big]
\cos\left(
\mathbf Q_{\rm CDW}\cdot\mathbf R+\theta_0
\right).\label{eq:SMDW}
\end{equation}
Consequently, thermal phase fluctuations suppress the density-wave coherence through an exponential renormalization factor, even when the electron-hole pairing amplitude remains finite.

The excitonic pairing amplitude vanishes at $T_g$, i.e., $|\Delta(T_g)|=0$,  determined from the self-consistent gap equation. In contrast, when thermal phase fluctuations become sufficiently strong, long-range density-wave order can disappear even while the excitonic pairing amplitude remains finite, with the transition temperature $T_c$ determined by
\begin{equation}
e^{-\langle\delta\theta^2(T_c)\rangle/2}\rightarrow 0,\quad 
\text{or equivalently}, \quad \langle\delta\theta^2(T_c)\rangle\rightarrow\infty.
\end{equation}
In the ideal case without pinning, the spontaneous breaking of continuous sliding symmetry here gives rise to a gapless Nambu-Goldstone mode (phason)~\cite{nambu1960quasi,nambu2009nobel,goldstone1961field,goldstone1962broken,littlewood81gauge}, whose thermal excitations generate phase fluctuations, and according to the Mermin-Wagner theorem~\cite{hohenberg1967existence,mermin1966absence,coleman1973there}, these fluctuations diverge in two dimensions at \emph{any} finite temperature, preventing true long-range density-wave order and yielding $T_c=0^+$. In realistic systems, however, lattice pinning explicitly breaks the continuous sliding symmetry and introduce a finite phason gap $m_\theta(T)$~\cite{PhysRevB.24.2751,gruner1988dynamics,gruner1985charge,maki1986thermal,PhysRevB.17.535,gpbp-qhp9}, thereby suppressing long-wavelength phase fluctuations and stabilizing long-range order at finite temperatures. If the effective pinning weakens sufficiently with increasing temperature such that $m_\theta(T)$ approaches zero, the phase fluctuations can diverge, leading to the disappearance of long-range density-wave order at a large finite $T_c$. Consequently, in the temperature range $T_c<T<T_g$, the system enters an excitonic pseudogap regime characterized by a finite single-particle gap but vanishing observable density-wave order. In this regime, strong thermal phase fluctuations destroy long-range phase coherence while electron-hole pairing amplitude persist, giving rise to phase-incoherent preformed excitons. This  pseudogap behavior originates from the distinct energy scales governing exciton-insulator pairing amplitude and phase coherence: the former is determined by the self-consistent gap equation and the underlying fermionic quasiparticle spectrum, whereas the latter is controlled by thermal fluctuations of the bosonic phason mode.

\subsection{Collective phase dynamics}

\subsubsection{Effective action for the phase-fluctuating finite-momentum excitonic insulator}

To evaluate the phase fluctuations of the intervalley excitonic condensate, we derive the effective action governing its collective phase dynamics. We consider the fermionic fields $\psi_{c,+}(x)$, $\psi_{c,-}(x)$, and $\psi_v(x)$, associated with the two conduction-band valleys centered at $\pm\mathbf Q_e$ and the valence band centered at the $\Gamma$ point, respectively, where $x=(\mathbf x,\tau)$. The interacting action reads
\begin{eqnarray}
S
\!=\!\int\!\!{dx}\Big\{
\sum_{\eta=\pm}
\psi_{c,\eta}^\dagger(x)
\left[
\partial_\tau\!+\!
\varepsilon_c(-i\nabla)\!-\!\mu
\right]
\psi_{c,\eta}(x)\!+\!
\psi_v^\dagger(x)
\left[
\partial_\tau\!+\!
\varepsilon_v(-i\nabla)\!-\!\mu
\right]
\psi_v(x)\!-\!
g\sum_{\eta=\pm}
\psi_{c,\eta}^\dagger(x)
\psi_v(x)
\psi_v^\dagger(x)
\psi_{c,\eta}(x)
\Big\}.~~~
\end{eqnarray}
In the presence of time-reversal symmetry, the two conduction-band valleys centered at $\pm\mathbf Q_e$ are related by time reversal. Accordingly, the excitonic pairing exhibits a valley-selective structure, with valence-band states in the $p_y>0$ ($p_y<0$) region predominantly coupled to the conduction valley at $+\mathbf Q_e$ ($-\mathbf Q_e$). We therefore decompose the valence-band fermionic field into right-moving and left-moving components,
\begin{equation}
\psi_v(x)=\psi_{v,+}(x)+\psi_{v,-}(x), \quad \text{where} \quad  \psi_{v,+}: p_y>0; \quad \psi_{v,-}: p_y<0.
\end{equation}
Hereafter, we focus on the $\eta=+$ pairing channel, with the $\eta=-$ channel obtained by time-reversal symmetry, consistent with the treatment in Sec.~\ref{SMham}, and then, we omit the valley index $\eta$ for simplicity. Accordingly, one can decouple the electron-hole interaction in the pairing channels by introducing the Hubbard-Stratonovich fields~\cite{schrieffer1964theory}
\begin{equation}
\Delta(x)
=
g
\left\langle
\psi_{c}^\dagger(x)
\psi_{v}(x)
\right\rangle.
\end{equation}
The resulting action becomes
\begin{align}
S=\int\!dx
\Bigg\{
&\Big(\psi^{\dagger}_{c}(x),\psi^{\dagger}_{v}(x)\Big)
\begin{pmatrix}
\partial_\tau+\varepsilon_c(-i\nabla)-\mu
&
-\Delta(x)
\\
-\Delta^*(x)
&
\partial_\tau+\varepsilon_v(-i\nabla)-\mu
\end{pmatrix}
\begin{pmatrix}
\psi_{c}(x)\\
\psi_{v}(x)
\end{pmatrix}
+\frac{|\Delta(x)|^2}{g}
\Bigg\}.
\end{align}
For the finite-momentum excitonic order, we express the order parameters as
\begin{align}
\Delta(x)
=
\Delta_{\bf q}e^{i(\mathbf Q_e+\mathbf q)\cdot\mathbf{x}}.
\end{align}
By performing a unitary transformation on the conduction-band fermionic fields, $\psi_{c}(x)
{\rightarrow}e^{i[(\mathbf Q_e+\mathbf q)\cdot\mathbf{x}]}
{\psi}_{c}(x)$, one has
\begin{align}
S=\int\!dx
\Bigg\{
&\Big({\psi}^{\dagger}_{c}(x),
{\psi}^{\dagger}_{v}(x)\Big)
\begin{pmatrix}
\partial_\tau+\xi_{c}(-i\nabla+\mathbf q)
&
-\Delta_{\bf q}
\\
-\Delta_{\bf q}^*
&
\partial_\tau+\xi_v(-i\nabla)
\end{pmatrix}
\begin{pmatrix}
{\psi}_{c}(x)\\
{\psi}_{v}(x)
\end{pmatrix}
+
\frac{|\Delta_{\bf q}|^2}{g}
\Bigg\}.
\end{align}
Here, we have redefined the band dispersions with momenta measured relative to the corresponding pocket centers,
\begin{align}
\xi_{c}(\mathbf k)
=\varepsilon_c(\mathbf k+\mathbf Q_e)-\mu,\qquad
\xi_v(\mathbf k)=\varepsilon_v(\mathbf k)-\mu.
\end{align}
This action then becomes the functional-integral representation of the mean-field Hamiltonian introduced in the main text.\\

The order parameter of an electronic condensate takes a generalized complex form, 
\begin{equation}
\Delta_{\mathbf q}(x)
=
\left[|\Delta_{\mathbf q}|+\delta|\Delta_{\mathbf q}(x)|\right]
e^{i[\theta_0+\delta\theta(x)]},
\end{equation}
where $\delta|\Delta_{\mathbf q}(x)|$ and $\delta\theta(x)$ describe the amplitude and phase dynamic fluctuations, respectively. In the following, we focus on phase fluctuations of the excitonic order parameter while neglecting amplitude fluctuations. Previous studies of fermionic condensates~\cite{PhysRevB.70.214531,PhysRevB.97.054510} have shown that the leading contributions from amplitude and short-wavelength phase fluctuations can largely cancel, resulting in relatively small corrections to the mean-field gap equation. Based on this cancellation mechanism, we retain the self-consistently determined pairing amplitude and focus on long-wavelength phase fluctuations (low-lying excitations). To derive the effective action for collective phase dynamics, a local unitary transformation is performed on the fermionic fields,
\begin{align}
\psi_{c}(x)\rightarrow
e^{i[\theta_0+\delta\theta(x)]/2}\psi_{c}(x),\qquad
\psi_{v}(x)\rightarrow
e^{-i[\theta_0+\delta\theta(x)]/2}\psi_{v}(x),
\end{align}
and the resulting action reads
\begin{align}
S=\int dx\,
\Bigg\{
&
\begin{pmatrix}
\psi_{c}^{\dagger}(x) &
\psi_{v}^{\dagger}(x)
\end{pmatrix}
\begin{pmatrix}
\partial_\tau+\frac{i}{2}\partial_\tau\delta\theta
+\xi_c\left(-i\nabla+\mathbf q+\mathbf{p}_s\right)
&
-|\Delta_{\mathbf q}|
\\
-|\Delta_{\mathbf q}|
&
\partial_\tau-\frac{i}{2}\partial_\tau\delta\theta
+\xi_v\left(-i\nabla-\frac12\nabla\delta\theta\right)
\end{pmatrix}
\begin{pmatrix}
\psi_{c}(x)\\
\psi_{v}(x)
\end{pmatrix}
+
\frac{|\Delta_{\mathbf q}|^2}{g}
\Bigg\},
\end{align}
where the gauge-invariant phase gradient $\mathbf{p}_s={\nabla\delta\theta}/{2}$. Integrating out the fermion fields generates the functional 
\begin{align}
S_{\rm eff}[\delta\theta]
=&\int dR\,\Big[\frac{|\Delta_{\mathbf q}|^2}{g}
-\mathrm{Tr}\ln(-G^{-1})\Big],
\label{EIaction}
\end{align}
where $\mathrm{Tr}$ denotes the trace over fermionic indices and space-time coordinates~\cite{peskin2018introduction}, and $G^{-1}$ is the inverse fermionic Green's function in the presence of phase fluctuations,
\begin{align}
G^{-1}(\mathbf{k},i\omega_n)
=&
i\omega_n
-
\xi_+(\mathbf{k},\mathbf{p}_s)\tau_0
-
\Big[
\xi_-(\mathbf{k},\mathbf{p}_s)
+
\frac{i\partial_\tau\delta\theta}{2}
\Big]\tau_3
-
|\Delta_{\bf q}|\tau_1,
\label{EIGreen_full}
\end{align}
with
\begin{align}
\xi_\pm(\mathbf{k},\mathbf{p}_s)
=
\frac{
\xi_c(\mathbf{k}+\mathbf q+\mathbf p_s)
\pm
\xi_v(\mathbf{k}-\mathbf p_s)
}{2}.
\end{align}
Here, $\tau_i$ are the Pauli matrices in the conduction-valence band space. Consistent with the preceding treatment, $\mathbf{k}$ is restricted to $k_y>0$, with the $k_y<0$ contribution obtained by time-reversal symmetry.

For isotropic parabolic electron and hole dispersions,
\begin{align}
\xi_c(\mathbf{k+q+p_s})
= 
\xi_c(\mathbf{k+q})
+
\frac{\mathbf{k+q}}{m_c}\cdot\mathbf{p}_s
+
\frac{p_s^2}{2m_c},\qquad
\xi_v(\mathbf{k-p_s})
=
\xi_v(\mathbf{k})
+
\frac{\mathbf{k}}{m_v}\cdot\mathbf{p}_s
-
\frac{p_s^2}{2m_v}.
\end{align}
and hence, 
\begin{align}
\xi_+(\mathbf{k},\mathbf{p}_s)
=&
\frac{
\xi_c(\mathbf{k+q})
+
\xi_v(\mathbf{k})
}{2}
+
\frac12
\left(
\frac{\mathbf{k+q}}{m_c}
+
\frac{\mathbf{k}}{m_v}
\right)
\cdot\mathbf{p}_s+
\frac{p_s^2}{4}
\left(
\frac1{m_c}
-
\frac1{m_v}
\right),
\label{Xiplus}\\
\xi_-(\mathbf{k},\mathbf{p}_s)
=&
\frac{
\xi_c(\mathbf{k+q})
-
\xi_v(\mathbf{k})
}{2}
+
\frac12
\left(
\frac{\mathbf{k+q}}{m_c}
-
\frac{\mathbf{k}}{m_v}
\right)
\cdot\mathbf{p}_s+
\frac{p_s^2}{4}
\left(
\frac1{m_c}
+
\frac1{m_v}
\right).
\label{Ximinus}
\end{align}
Expanding the effective action in powers of the phase fluctuations yields
\begin{align}
S_{\rm eff}
=
S_0
+
S^{(2)}
+\cdots .
\end{align}
The zeroth-order effective action is
\begin{align}
S_0
=
\int dR
\left[
\frac{|\Delta|^2}{g}
-
{\rm Tr}\ln(-G_0^{-1})
\right],
\end{align}
where the equilibrium inverse Green's function is 
$G_0^{-1}
=
i\omega_n
-
\xi_+\tau_0
-
\xi_-\tau_3
-
\Delta\tau_1$,
with the corresponding Green's function~\cite{abrikosov2012methods,mahan2013many}
\begin{align}
G_0
=
\frac{
(i\omega_n-\xi_+)\tau_0
+
\xi_-\tau_3
+
\Delta\tau_1
}{
(i\omega_n-E_{\bf k}^+)(i\omega_n-E_{\bf k}^-)
}.
\label{G0}
\end{align}
Here, $E_{\mathbf{k}}^\pm
=
\xi_+\pm E_{\mathbf{k}}$ with 
$E_{\mathbf{k}}
=
\sqrt{\xi_-^2+|\Delta_{\mathbf q}|^2}$.\\

As a self-consistent check, the saddle-point condition 
$\frac{\partial S_0}{\partial |\Delta|}
=0$ 
gives
\begin{align}
\frac{1}{g}
=-
T\sum_{\mathbf{k},\omega_n}
\frac{1}
{
(i\omega_n-\xi_+)^2-E_{\mathbf{k}}^2
}=
\sum_{\mathbf{k}}
\frac{
f(E_{\mathbf{k}}^-)
-
f(E_{\mathbf{k}}^+)
}
{2E_{\mathbf{k}}},
\end{align}
and we obtain the exactly same gap equation as the one in Eq.~(\ref{eq:SM_gap}).\\

The quadratic contribution to the effective action describes the collective phase dynamics, and is given by
\begin{align}
S^{(2)}
=
\int dR
\Big[
{\rm Tr}(G_0\Sigma^{(2)})
+
\frac12
{\rm Tr}
(G_0\Sigma^{(1)}G_0\Sigma^{(1)})
\Big],
\end{align}
where $\Sigma^{(1)}$ and $\Sigma^{(2)}$ denote the first- and second-order corrections to the inverse Green's function arising from phase fluctuations.

The contribution from the second-order self-energy correction, corresponding to the diamagnetic response~\cite{mahan2013many}, is
\begin{align}
{\rm Tr}
\Big(
G_0\Sigma^{(2)}
\Big)
=
\frac{p_s^2}{4}
\Big[
\Big(
\frac{1}{m_c}
-
\frac{1}{m_v}
\Big)
\chi_0
+
\Big(
\frac{1}{m_c}
+
\frac{1}{m_v}
\Big)
\chi_3
\Big],
\label{SEI1chi}
\end{align}
where 
\begin{align}
\chi_0
&=
{\rm Tr}(G_0\tau_0)=
\sum_{\mathbf{k},\omega_n}
\frac{
2(i\omega_n-\xi_+)
}
{
(i\omega_n-\xi_+)^2-E_{\bf k}^2
}=
\sum_{\mathbf{k}}
\left[
f(E_{\mathbf{k}}^+)
+
f(E_{\mathbf{k}}^-)
-1
\right],
\label{chi0}\\
\chi_3&=
{\rm Tr}(G_0\tau_3)=
\sum_{\mathbf{k},\omega_n}
\frac{
2\xi_-
}
{
(i\omega_n-\xi_+)^2-E_{\bf k}^2
}=
\sum_{\mathbf{k}}
\frac{\xi_-}{E_{\mathbf{k}}}
\left[
f(E_{\mathbf{k}}^+)
-
f(E_{\mathbf{k}}^-)
\right].
\label{chi3}
\end{align}
The contribution quadratic in the first-order self-energy correction gives the paramagnetic response~\cite{mahan2013many},
\begin{align}
\frac{1}{2}
{\rm Tr}
\left(
G_0\Sigma^{(1)}
G_0\Sigma^{(1)}
\right)=-\frac{1}{2}
\left(\frac{\partial_\tau\delta\theta}{2}\right)^2
\chi_{33}+
\frac{1}{2}
p_{s,i}p_{s,j}
\Big[
v_{+,i}v_{+,j}\chi_{00}
+
v_{+,i}v_{-,j}\chi_{03}
+
v_{-,i}v_{+,j}\chi_{30}
+
v_{-,i}v_{-,j}\chi_{33}
\Big].
\label{Sparachi}
\end{align}
with the velocity combinations
\begin{align}
v_{+,i}
=
\frac{1}{2}
\left(
\frac{k_i+q_i}{m_c}
+
\frac{k_i}{m_v}
\right),\qquad
v_{-,i}=
\frac{1}{2}
\left(
\frac{k_i+q_i}{m_c}
-
\frac{k_i}{m_v}
\right).
\end{align}
as well as the response functions 
\begin{align}
\chi_{03}
&=
T\sum_{\omega_n,\mathbf{k}}
{\rm Tr}
\left(
G_0\tau_0G_0\tau_3
\right)=
4T\sum_{\omega_n,\mathbf{k}}
\frac{
(i\omega_n-\xi_+)\xi_-
}
{
\left[
(i\omega_n-\xi_+)^2-E_{\mathbf{k}}^2
\right]^2
}=\sum_{\mathbf{k}}
\frac{\xi_-\left[
f'(E_{\mathbf{k}}^+)
-
f'(E_{\mathbf{k}}^-)
\right]}{E_{\mathbf{k}}},\\
\chi_{30}&=\chi_{03},\\
\chi_{00}
&=
T\sum_{\omega_n,\mathbf{k}}
{\rm Tr}
\left(
G_0\tau_0G_0\tau_0
\right)
=
2T\sum_{\omega_n,\mathbf{k}}
\frac{
(i\omega_n-\xi_+)^2+E_{\mathbf{k}}^2
}
{
\left[
(i\omega_n-\xi_+)^2-E_{\mathbf{k}}^2
\right]^2
}=
\sum_{\mathbf{k}}
\left[
f'(E_{\mathbf{k}}^+)
+
f'(E_{\mathbf{k}}^-)
\right],\\
\chi_{33}
&=
T\sum_{\omega_n,\mathbf{k}}
{\rm Tr}
\left(
G_0\tau_3G_0\tau_3
\right)=
2T\sum_{\omega_n,\mathbf{k}}
\frac{
(i\omega_n-\xi_+)^2+\xi_-^2-\Delta^2
}
{
\left[
(i\omega_n-\xi_+)^2-E_{\mathbf{k}}^2
\right]^2
}=
\chi_{00}
-
4\Delta^2
T\sum_{\omega_n,\mathbf{k}}
\frac{1}
{
\left[
(i\omega_n-\xi_+)^2-E_{\mathbf{k}}^2
\right]^2}\nonumber\\
&=
\sum_{\mathbf{k}}
\Big[
\frac{\xi_-^2}{E_{\mathbf{k}}^2}
\left(
f'(E_{\mathbf{k}}^+)
+
f'(E_{\mathbf{k}}^-)
\right)
+
\frac{|\Delta|^2}{E_{\mathbf{k}}^3}
\left(
f(E_{\mathbf{k}}^+)
-
f(E_{\mathbf{k}}^-)
\right)
\Big].
\label{chi33def}
\end{align}
Consequently, the derived quadratic  action of phase dynamics can be written as
\begin{align}
S^{(2)}
=\int{dR}\Big[\frac18
D(\partial_\tau\delta\theta)^2
+
\frac12
p_{s,i}p_{s,j}f^c_{ij}\Big],
\label{Sstiffness}
\end{align}
where the phase stiffness tensor is
\begin{align}
f^c_{ij}
=
\frac{\delta_{ij}}{2}
\left[
\left(
\frac{1}{m_c}
-
\frac{1}{m_v}
\right)
\chi_0
+
\left(
\frac{1}{m_c}
+
\frac{1}{m_v}
\right)
\chi_3
\right]
+
v_{+,i}v_{+,j}\chi_{00}
+
v_{+,i}v_{-,j}\chi_{03}
+
v_{-,i}v_{+,j}\chi_{30}
+
v_{-,i}v_{-,j}\chi_{33}.
\label{rhoij}
\end{align}
and $D=-\chi_{33}$.

\subsubsection{Phase pinning induced by lattice commensurability}

In realistic density-wave systems, the phase mode can acquire an excitation
gap due to the mismatch between the density-wave modulation and the underlying
lattice periodicity~\cite{PhysRevB.24.2751,gruner1988dynamics,gruner1985charge,maki1986thermal,PhysRevB.17.535,gpbp-qhp9}. This lattice-induced pinning provides an effective mass
for the phason mode and regularizes the infrared divergence of phase
fluctuations. Specifically, as derived above, the phase fluctuation dynamics of the excitonic density-wave state is described
by the effective action
\begin{align}
S^{(2)}
=
\frac18
\int dR
\Big[
D(\partial_\tau\delta\theta)^2
+
f^c_{ij}
(\partial_i\delta\theta)
(\partial_j\delta\theta)
\Big],
\label{Sphase}
\end{align}
where $D$ is the temporal phase stiffness and $f^c_{ij}$ is the spatial
stiffness tensor obtained from the microscopic fermionic response. The coupling between the density-wave order and the lattice commensurability can be
written as~\cite{PhysRevB.17.535,maki1986thermal}
\begin{align}
S_{\rm pin}
=
-\lambda
\int dR
\cos
\big[{l}(
\delta{\bf Q}\cdot{\bf R}+\theta_0
+\delta\theta(R))
\big],
\label{Spinlat}
\end{align}
where $\delta{\bf Q}={\bf Q}_{\rm CDW}-{\bf Q}_{\rm lat}$ denotes the wave-vector mismatch between the density-wave modulation and the relevant Fourier component of the lattice potential (the lattice-potential wave vector ${\bf Q}_{\rm lat}$ is generally a rational fraction of a reciprocal lattice vector, i.e., ${\bf Q}_{\rm lat}={\bf G}/l$, where $l$ is an integer), and $\lambda$ characterizes the
lattice-induced pinning strength~\cite{PhysRevB.17.535,maki1986thermal}. Our calculations identify ${\bf Q}_{\rm CDW}\approx(0,2\pi/5b)$, indicating that the strongest lattice-induced pinning occurs for $l=5$, thereby yielding $\delta{\bf Q}\approx0$.

The equation of motion for the phase fluctuation is obtained from the
Euler-Lagrange equation~\cite{peskin2018introduction},
\begin{align}
{\partial_\mu
\left[
\partial(S^{(2)}+S_{\rm pin})
/
\partial(\partial_\mu\delta\theta)
\right]}
=
\frac{\partial(S^{(2)}+S_{\rm pin})}
{\partial(\delta\theta)},
\end{align}
which gives
\begin{align}
\left[
\frac{D}{4}\partial_t^2
-
\frac14f^c_{ij}\partial_i\partial_j
\right]
\delta\theta(R)
+
l\lambda
\sin
\left[{l}(
\delta{\bf Q}\cdot{\bf R}+\theta_0
+\delta\theta(R))
\right]
=0,\qquad\text{with}\quad \delta{\bf Q}\approx0.
\label{EOM0}
\end{align}
To incorporate the effect of phase fluctuations, we employ a cumulant
expansion and retain only one uncontracted phase field, while contracting the
remaining fields according to Wick's theorem~\cite{peskin2018introduction}. Using 
$\sin(l\theta_e+l\delta\theta)
=
\sin{l}\theta_e\cos{l}\delta\theta
+
\cos{l}\theta_e\sin{l}\delta\theta$ and assuming 
$\langle\delta\theta\rangle=0$, the linear response contribution is determined by the second term. The sine term is expanded as
\begin{align}
\sin(l\delta\theta)
=
\sum_{n=0}^{\infty}
\frac{(-1)^n}{(2n+1)!}
(l\delta\theta)^{2n+1}.
\end{align}
Within the cumulant approximation, $(l\delta\theta)^{2n+1}
\rightarrow
(2n-1)!!
\langle(l\delta\theta)^2\rangle^n
l\delta\theta$, which gives $\sin(l\delta\theta)
\rightarrow
\exp[
-\frac12
\langle(l\delta\theta)^2\rangle]l\delta\theta$. Therefore,
\begin{align}
\sin[l(\theta_e+\delta\theta)]
\rightarrow
\cos(l\theta_e)
\exp
\Big[
-\frac{l^2}{2}
\langle(\delta\theta)^2\rangle
\Big]
l\delta\theta .
\end{align}
Substituting this result into Eq.~(\ref{EOM0}), one has
\begin{align}
\Big[
\frac{D}{4}\partial_t^2
-
\frac14f^c_{ij}\partial_i\partial_j
+
l^2\lambda
\cos\theta_e
\exp\Big(
-\frac{l^2}{2}
\langle(\delta\theta)^2\rangle
\Big)
\Big]
\delta\theta 
=0.
\label{EOM$_1$}
\end{align}
For equilibrium with $\delta{\bf Q}\approx0$ and  
$\langle\cos({l}\theta_e)\rangle=\langle\cos[{l}(\delta{\bf Q}\cdot{\bf R}+\theta_0)]\rangle\approx\langle\cos(l\theta_0)\rangle\approx1$~\cite{PhysRevB.17.535}, the phase equation becomes 
\begin{align}
\Big[
D\partial_t^2
-
f^c_{ij}\partial_i\partial_j
+
4l^2\lambda
\exp
\Big(
-\frac{l^2}{2}
\langle(\delta\theta)^2\rangle
\Big)
\Big]
\frac{\delta\theta(R)}{2}
=0 .
\label{EOM$_2$}
\end{align}
Consequently, the phason excitation spectrum is~\cite{PhysRevB.24.2751,maki1986thermal,gpbp-qhp9}
\begin{align}
\omega^2({\bf q})
=
m_{\theta}^2(T)
+
\frac{f^c_{ij}}{D}q_iq_j ,
\label{phasonspectrum}
\end{align}
where the temperature-dependent phason mass is
\begin{align}
m_{\theta}^2(T)
=
\frac{4l^2\lambda}{D}
\exp
\left[
-\frac{l^2}{2}
\langle(\delta\theta)^2\rangle
\right]=m_{\theta}^2(T=0)
\exp\Big[
-\frac{l^2}{2}
\langle(\delta\theta)^2\rangle\Big].
\label{phasonmass}
\end{align}
The phase fluctuation therefore renormalizes the lattice-induced pinning
strength and continuously suppresses the phason excitation gap with increasing temperature, consistent with previous theoretical  studies~\cite{PhysRevB.24.2751,maki1986thermal,gpbp-qhp9}.

\begin{figure}[h]
\centering
\includegraphics[width=0.9\linewidth]{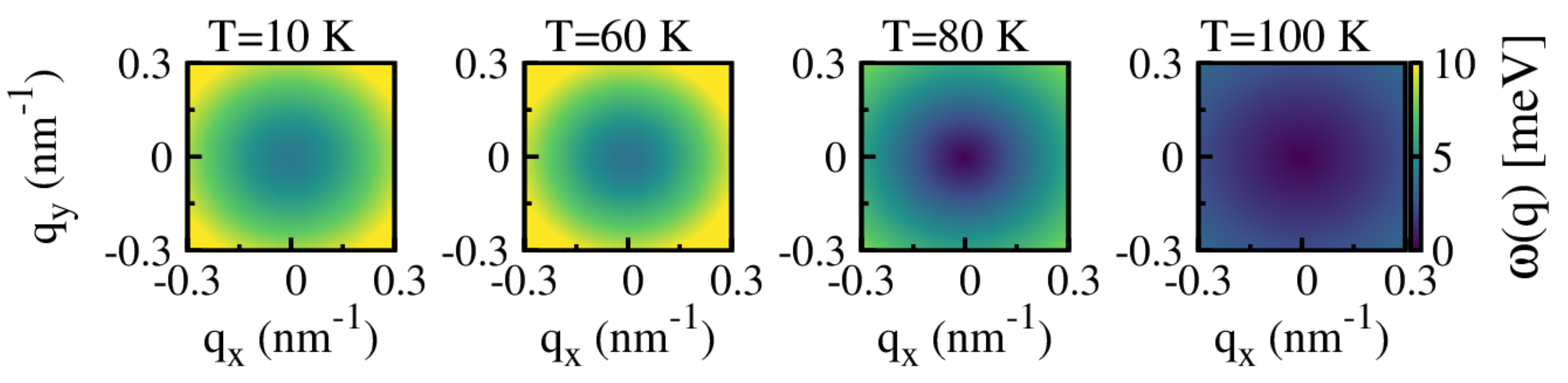}
\caption{Calculated phason dispersion at different temperatures.}
\label{SFIG1}
\end{figure}

\subsubsection{Thermal phase fluctuations}

Following the above derivation, the Gaussian effective action governing the collective phase fluctuations can be expressed as
\begin{align}
S^{(2)}
=
\frac12
\sum_{\mathbf q,\Omega_n}
\Big[
\frac{D}{4}
(\Omega_n^2+m_{\theta}^2)
+
\frac{f^c_{ij}}{4}q_iq_j\Big]
|\delta\theta(\mathbf q,\Omega_n)|^2,
\end{align}
and then, the corresponding phase correlation function is~\cite{peskin2018introduction,abrikosov2012methods,mahan2013many}
\begin{align}
\langle
\delta\theta(\mathbf q,\Omega_n)
\delta\theta(-\mathbf q,-\Omega_n)\rangle
=
\frac{4}
{
D(\Omega_n^2+m_{\theta}^2)
+
f^c_{ij}q_iq_j
}.
\end{align}
Accordingly, the mean-square phase fluctuations are given by
\begin{align}
\Big\langle
\Big(\frac{\delta\theta}{2}\Big)^2
\Big\rangle
=
T
\sum_{\Omega_n}
\int
\frac{d^2q}{(2\pi)^2}
\frac{1}
{
D(\Omega_n^2+m_{\theta}^2)
+
f^c_{ij}q_iq_j
}.
\end{align}
Performing the bosonic Matsubara frequency summation yields
\begin{align}
\Big\langle
\left(\frac{\delta\theta}{2}\right)^2
\Big\rangle
=
\frac{1}{2D}
\int
\frac{d^2q}{(2\pi)^2}
\frac{
1+2n_B(\omega_{\mathbf q})
}
{\omega_{\mathbf q}},
\end{align}
where $n_B(\omega)$ is the Bose-Einstein distribution function, and the phason dispersion is 
$\omega_{\mathbf q}
=
\sqrt{
m_{\theta}^2
+{f^c_{ij}q_iq_j}/{D}
}$. The phase fluctuations contain both zero-point and thermal contributions, corresponding to the two terms in the numerator. When the model parameters are determined from zero-temperature experimental measurements, the effects of zero-point fluctuations are typically already incorporated into the renormalized ground-state parameters~\cite{peskin2018introduction,gpbp-qhp9,yang2024first}. We therefore retain only the thermal phase fluctuations in the following analysis. The thermal phase fluctuations are thus evaluated as
\begin{align}
\Big\langle
\Big(\frac{\delta\theta}{2}\Big)^2
\Big\rangle_{\rm th}
=
\frac{1}{D}
\int
\frac{d^2q}{(2\pi)^2}
\frac{
n_B(\omega_{\mathbf q})
}
{\omega_{\mathbf q}}.\label{eq:thph}
\end{align}

As shown in Fig.~\ref{SFIG1}, as the temperature increases, the phason undergoes two self-consistently coupled softening mechanisms. First, the excitation gap ($m_{\theta}$) at $\mathbf q=0$ progressively decreases due to thermal phase fluctuations, becoming nearly gapless around 80~K. Second, the phason velocity ($c_{ii}=\sqrt{f^{c}_{ii}/D}$) gradually decreases due to the thermal suppression of the phase stiffness. Eventually, the phason dispersion collapses near 104~K, where the excitonic pairing amplitude $|\Delta(T)|$ vanishes.

{\sl Note added.---}Compactness and Gaussianity describe two distinct aspects of the phase field~\cite{PhysRevB.24.2751,maki1986thermal,yang2026superconducting}: the former specifies its physical configuration space $S^1$, while the latter defines a Gaussian trial measure on its universal covering space $\mathbb{R}$. Within a smooth topological sector, one can write $\theta(x)=\theta_0+\delta\theta(x)\ \mathrm{mod}\ 2\pi$, where $\delta\theta\in\mathbb{R}$ follows a Gaussian distribution with variance $\sigma^2=\langle\delta\theta^2\rangle_G$. The corresponding probability density on the compact circle is obtained by wrapping the Gaussian distribution,
\begin{equation}
P_{S^1}(\theta)=\frac{1}{\sqrt{2\pi\sigma^2}}
\sum_{m\in\mathbb{Z}}
\exp\left[-\frac{(\theta-\theta_0+2\pi m)^2}{2\sigma^2}\right].
\end{equation}
This distribution is periodic and normalized for any finite $\sigma^2>0$, without requiring small phase fluctuations. In particular, periodic observables can be evaluated exactly using the Gaussian characteristic function,
\begin{equation}
\left\langle e^{in\theta}\right\rangle
=e^{in\theta_0}e^{-n^2\sigma^2/2},
\qquad
\left\langle\cos(n\theta)\right\rangle
=\cos(n\theta_0)e^{-n^2\sigma^2/2},\quad \text{where $n\in\mathbb{Z}$}.
\end{equation}
Thus, the exponential factor is a consequence of Gaussian statistics rather than a small-fluctuation expansion. As $\sigma^2\rightarrow\infty$, the wrapped Gaussian approaches the uniform distribution $P_{S^1}(\theta)=1/(2\pi)$, and all nonzero harmonics vanish. Therefore, Gaussian phase fluctuations remain mathematically compatible with compactness even when their variance becomes large.

\begin{figure}[h]
\centering
\includegraphics[width=0.3\linewidth]{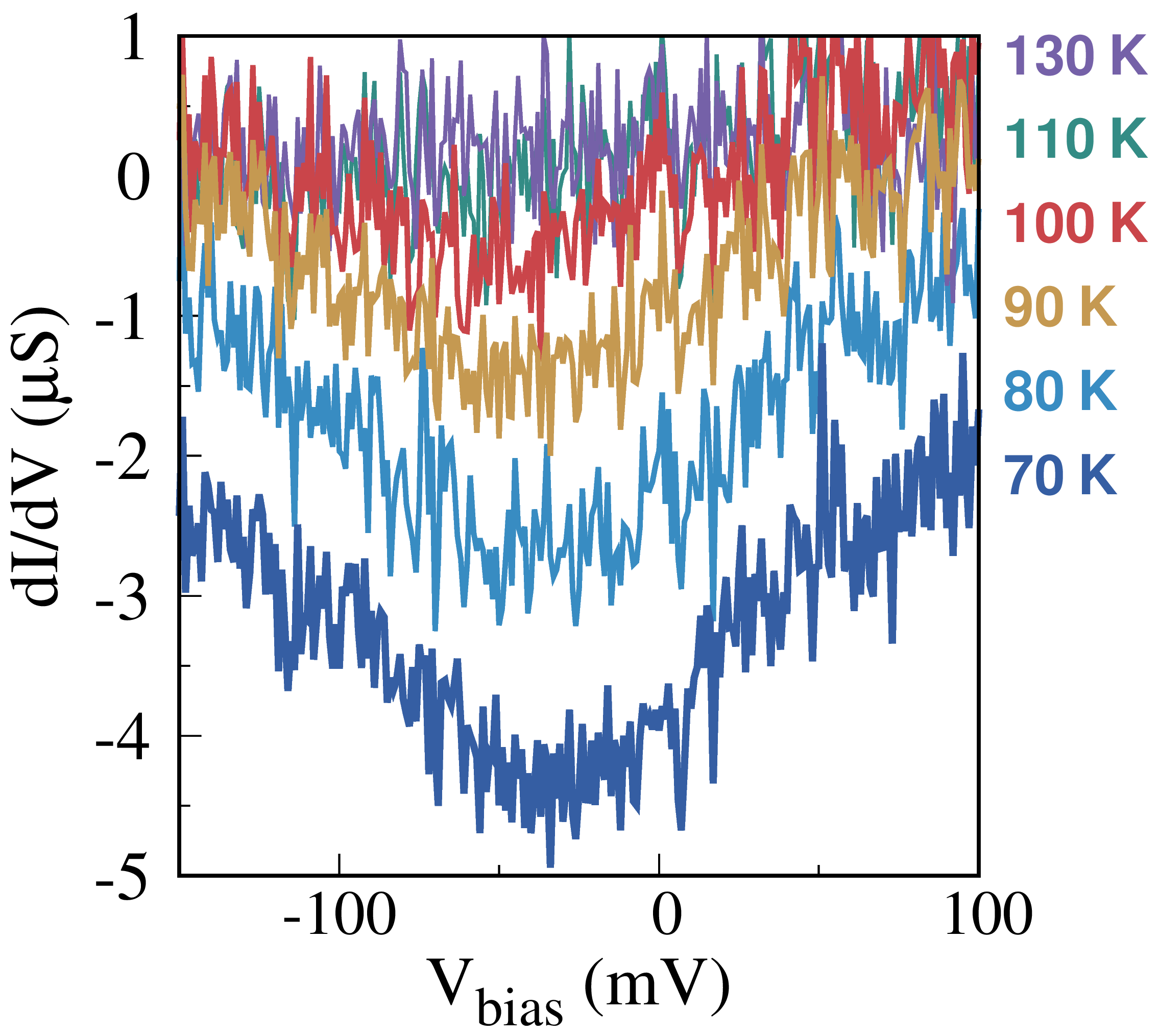}
\caption{Temperature-dependent differential tunneling conductance after subtraction of the normal-state spectrum at 140~K, $(\mathrm{d}I/\mathrm{d}V)_T-(\mathrm{d}I/\mathrm{d}V)_{140\,\mathrm{K}}$, corresponding to the inset of Fig.~2(a) of the main text.}
\label{SFIG4}
\end{figure}


\section{Additional indirect experimental indication for the gap-opening temperature}

\begin{figure}[h]
\centering
\includegraphics[width=\linewidth]{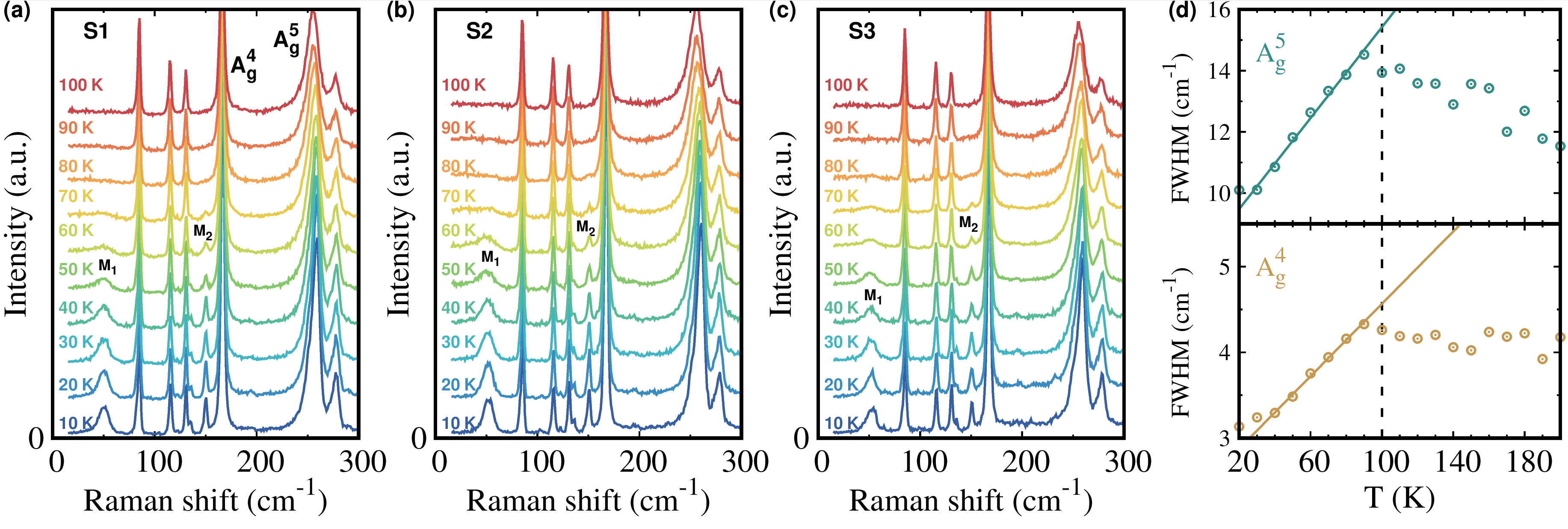}
\caption{(a)-(c) Raman spectra measured in the $c(bb)\bar{c}$ geometry for three samples (S1, S2, and S3), respectively.
(d) Temperature dependence of the full width at half maximum (FWHM) of the $A_g^{4}$ and $A_g^{5}$ phonon modes. The solid lines represent linear fits to the data over the temperature range of 30-90 K, capturing the linewidth evolution accompanying the closing of the single-particle gap. The data points above 100~K deviate from this linear behavior.}
\label{SFIG2}
\end{figure}

In addition to the resistivity and tunneling spectroscopy measurements [see Fig.~\ref{SFIG4}] presented in the main text, we provide an independent indication of the single-particle gap-opening temperature by examining the temperature dependence of ordinary phonon linewidths. Upon excitonic-insulator formation, electron-hole pairing gaps out electronic states in the overlapping regions of the electron and hole Fermi surfaces, reducing the available low-energy electronic excitations and consequently suppressing electron-phonon scattering. This modification of the electronic scattering phase space can be reflected in the linewidths of ordinary phonon modes. To investigate this effect, we select two representative phonon modes, $A_g^{4}$ and $A_g^{5}$, as shown in Fig.~\ref{SFIG2}(a), and track their full widths at half maximum (FWHM) as a function of temperature.

As shown in Fig.~\ref{SFIG2}(d), both phonon modes exhibit an approximately linear temperature dependence of their linewidths over 30-90 K. Above 100 K, the data clearly deviate from the linear extrapolation of this low-temperature behavior, indicating a crossover in the temperature evolution of the phonon linewidths around 90-100~K. The opening of the electronic gap reconstructs the low-energy electron-hole excitation spectrum, thereby modifying the available scattering channels and the associated phonon damping below 100~K, consistent with the resistivity and tunneling spectroscopy measurements.  Similar effects have also been reported in other CDW materials, such as EuAl$_4$~\cite{Cao2025EuAl4}, where the CDW transition is accompanied by a pronounced modification of the electron-phonon coupling and consequently modifies the phonon linewidths~\cite{Cao2025EuAl4}. 

\section{Analysis of M$_2$ mode}

\begin{figure}[h]
\centering
\includegraphics[width=\linewidth]{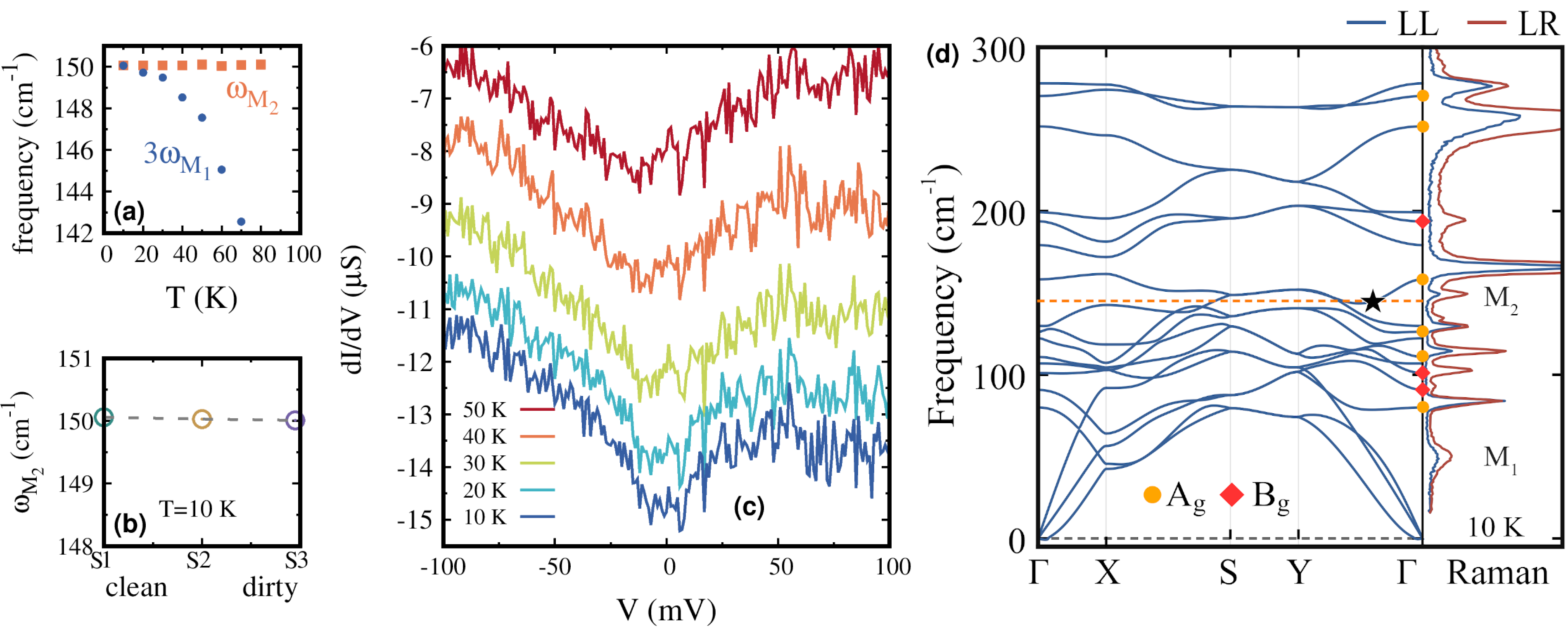}
\caption{(a) Temperature dependence of the $M_2$ mode frequency $\omega_{M_2}$, with the $M_1$ mode shown for comparison. (b) $M_2$ mode frequencies measured at 10 K for samples S1, S2, and S3. In contrast to $M_1$, the $M_2$ mode exhibits negligible temperature and disorder dependence. (c) Low-temperature tunneling conductance after subtraction of the normal-state spectrum at 140~K, $(\mathrm{d}I/\mathrm{d}V)_T-(\mathrm{d}I/\mathrm{d}V)_{140\,\mathrm{K}}$. The partially gapped Fermi surface leaves residual spectral weight within the gap, resulting in a non-ideal U-shaped gap profile without particle-hole symmetry. (d) DFT-calculated phonon band structure compared with polarization-resolved Raman spectra measured at 10~K. Red and orange symbols at the $\Gamma$ point indicate the Raman-active $B_g$ and $A_g$ phonon modes, respectively, while the black star marks the $5b$-folded $A_g^4$ phonon mode. The orange horizontal dashed line marks the energy of the  $5b$-folded $A_g^4$ phonon mode, which nearly coincides with the $M_2$ energy.}
\label{SFIG3}
\end{figure}

In contrast to the $M_1$ mode discussed extensively in the main text, which exhibits clear softening with increasing temperature and hardening with increasing disorder, the frequency of the $M_2$ mode remains nearly unchanged under either condition, as shown in Fig.~\ref{SFIG3}(a) and (b). This distinct behavior suggests that $M_2$ originates from a zone-folded phonon, consistent with the emergence of folded phonon modes commonly observed in CDW systems~\cite{PhysRevLett.122.127001,Liu2022}.  Notably, although the density-wave order in our system originates from electron-hole pairing, the resulting periodic modulation of the electronic density can induce a secondary weak lattice distortion through electron-phonon coupling. 
 
To illustrate this mechanism, we consider the minimal Hamiltonian
\begin{equation}
H=H_{\mathrm{EI}}+H_{\mathrm{ph}}+H_{\mathrm{ep}},
\end{equation}
where $H_{\mathrm{EI}}$ describes the electronic system supporting the excitonic condensate. The phonon Hamiltonian is given by
\begin{equation}
H_{\mathrm{ph}}
=
\sum_{\mathbf q}
\omega_{\mathbf q}
b_{\mathbf q}^{\dagger}b_{\mathbf q},
\end{equation}
where $b_{\mathbf q}^{\dagger}$ and $b_{\mathbf q}$ are the phonon creation and annihilation operators, respectively. The interband electron-phonon coupling reads
\begin{equation}
H_{\mathrm{ep}}
=
\frac{\gamma}{\sqrt N}
\sum_{\mathbf k,\mathbf q}(
c_{\mathbf k+\mathbf q}^{\dagger}v_{\mathbf k}
+\mathrm{h.c.})(
b_{\mathbf q}+b_{-\mathbf q}^{\dagger}),
\end{equation}
where $c_{\mathbf k}^{\dagger}$ and $v_{\mathbf k}$ denote the electron creation and annihilation operators in the conduction and valence bands, respectively, and $\gamma$ is the interband electron-phonon coupling strength. The finite-momentum excitonic condensate is characterized by 
\begin{equation}
\Phi_{\mathbf Q}
=
\frac{1}{N}
\sum_{\mathbf k}
\langle
c_{\mathbf k+\mathbf Q}^{\dagger}v_{\mathbf k}\rangle,
\end{equation}
where $\mathbf Q$ is the excitonic density-wave wavevector. Within the mean-field framework, the electron-phonon coupling generates a linear driving term for the phonon mode at $\mathbf Q$. The corresponding effective phonon Hamiltonian can be written as
\begin{equation}
H_{\mathrm{ph}}^{\mathrm{MF}}
=
\omega_{\mathbf Q}b_{\mathbf Q}^{\dagger}b_{\mathbf Q}
+
\gamma\sqrt N(
\Phi_{\mathbf Q}b_{\mathbf Q}^{\dagger}
+
\Phi_{\mathbf Q}^{*}b_{\mathbf Q}).
\end{equation}
Introducing the displaced phonon operator
\begin{equation}
\widetilde b_{\mathbf Q}
=
b_{\mathbf Q}
+
\frac{\gamma\sqrt N}{\omega_{\mathbf Q}}\Phi_{\mathbf Q},
\end{equation}
the effective Hamiltonian becomes
\begin{equation}
H_{\mathrm{ph}}^{\mathrm{MF}}
=
\omega_{\mathbf Q}
\widetilde b_{\mathbf Q}^{\dagger}\widetilde b_{\mathbf Q}
-
\frac{Ng^2}{\omega_{\mathbf Q}}|\Phi_{\mathbf Q}|^2.
\end{equation}
The equilibrium phonon displacement is therefore
\begin{equation}
\langle b_{\mathbf Q}\rangle
=
-\frac{h\sqrt N}{\omega_{\mathbf Q}}\Phi_{\mathbf Q},
\end{equation}
demonstrating that the excitonic condensate induces a static lattice distortion at the same ordering wavevector~\cite{gruner1985charge,gruner1988dynamics}. This distortion breaks the original translational symmetry, allowing finite-momentum phonons to be folded to the reduced Brillouin-zone center and become Raman active. Notably, the displacement transformation shifts only the equilibrium position of the phonon oscillator, leaving its harmonic excitation frequency unchanged, $\widetilde{\omega}_{\mathbf Q}=\omega_{\mathbf Q}$. This naturally accounts for the weak temperature and disorder dependence of the $M_2$ frequency, and distinguishes $M_2$ from the $M_1$ mode. 

In particular, for the experimentally observed $5b$ modulation, the corresponding folding wavevector is $\mathbf Q=2\pi/(5b)$. Notably, as shown in Fig.~\ref{SFIG3}(d), the frequency of the $A_g^{4}$ phonon branch at this wavevector, obtained from our DFT phonon calculations, agrees with our experimentally observed $M_2$ frequency,
\begin{equation}
\omega_{M_2}^{\mathrm{exp}}
\simeq
\omega_{A_g^{4}}^{\mathrm{DFT}}
\left(q_y=\frac{2\pi}{5b}\right),
\end{equation}
providing further support for the identification of $M_2$ as a zone-folded phonon.\\

The $M_1$ mode exhibits thermal softening [Fig.~2(c)] and systematic disorder-induced hardening [Fig.~2(d)], suggesting that $M_1$ is unlikely to originate from phonon folding but arises from the collective excitation of the condensate that is sensitive to both temperature and disorder.  Meanwhile, several considerations suggest against identifying $M_1$ as an amplitudon of the excitonic condensate. On the BCS side of the excitonic-insulator regime, the amplitude mode is expected to occur at a relatively high energy scale, typically comparable to the single-particle excitation threshold ($\omega_A\sim 2|\Delta(0)|$), as in conventional Higgs-mode theories~\cite{Pekker2015,PhysRevB.101.195118,Werdehausen2018,Shimano2020higgs,Chu2020,PhysRevB.98.094507,PhysRevB.100.104513,PhysRevB.102.014511}. In the present system, this energy scale is estimated to be approximately $38$ meV, substantially higher than the observed $\omega_{M_1}$. Moreover, in  conventional condensate-based descriptions~\cite{Pekker2015,PhysRevB.101.195118,Werdehausen2018,Shimano2020higgs,Chu2020,PhysRevB.98.094507,PhysRevB.100.104513,PhysRevB.102.014511}, the amplitudon typically exhibits a continuous thermal softening that becomes increasingly pronounced upon approaching the transition temperature $T_g$, reflecting the progressive suppression of the condensate amplitude. The observed temperature evolution of $M_1$ is distinct from such a mean-field gap behavior.  On the other hand, the $M_1$ mode exhibits pronounced hardening with increasing disorder, whereas the amplitudon would generally be expected to soften as disorder suppresses the excitonic condensate. These suggest that the $M_1$ mode is unlikely to be an amplitudon of the excitonic condensate.

\begin{figure}[h]
\centering
\includegraphics[width=0.5\linewidth]{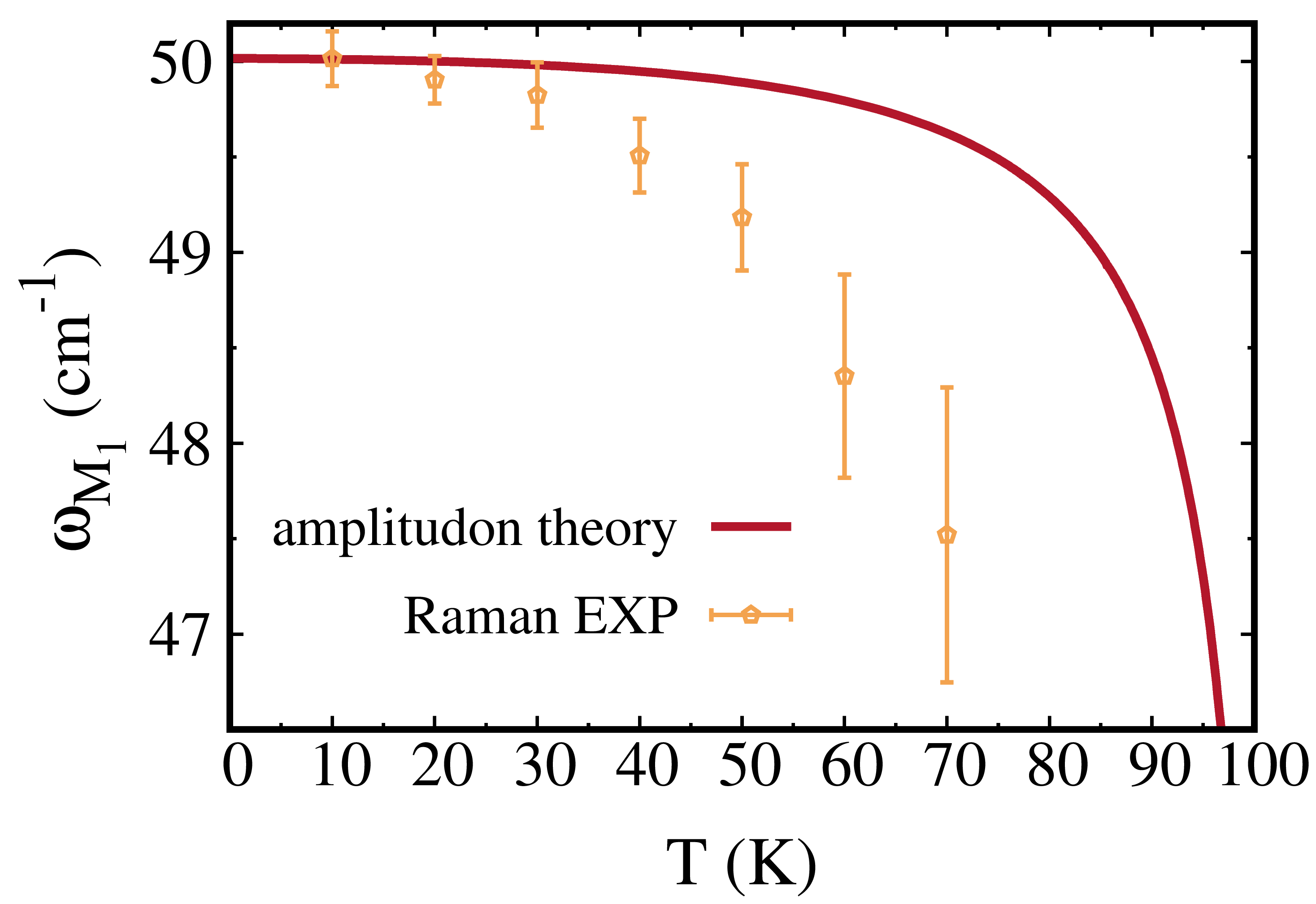}
\caption{
Comparison between the experimentally observed
$M_1$ Raman mode and the amplitudon predicted
for a lattice-driven CDW. The theoretical
amplitudon frequency, calculated from
Eq.~(\ref{amplitudon}), exhibits pronounced
softening as the CDW gap closes near 100~K,
in contrast to the experimentally observed
temperature dependence of the $M_1$ mode.
}
\label{AMFIG}
\end{figure}

\begin{itemize}
\item The present manuscript assumes an electronically driven excitonic insulator, since the intrinsic \emph{attractive} electron-hole
attraction provides the dominant instability, particularly in the presence of a substantial overlap between the conduction and valence bands, and our DFT phonon calculations reveal neither imaginary
phonon frequencies nor low-energy soft modes, indicating
the dynamical stability of the lattice. The density-wave order here emerges as a secondary consequence of the electronically driven EI~\cite{PhysRev.158.462,PhysRevLett.19.439,RevModPhys.40.755}, whose collective excitations originate from the electronic EI condensate itself.  

One issue concerns whether the observed $M_1$ mode can be attributed to the amplitudon of a lattice-driven CDW~\cite{gruner1985charge,gruner1988dynamics,gpbp-qhp9,lee1974conductivity}. Specifically, within a conventional lattice-driven CDW scenario, the amplitudon originates from collective amplitude
oscillations of the lattice distortion, with its characteristic frequency given by~\cite{gruner1985charge,gruner1988dynamics,gpbp-qhp9,lee1974conductivity} 
\begin{equation}\label{amplitudon}
\Omega^2_{\rm A}(T)
=\frac{\lambda\omega^2_{\bf Q}}
{1+\lambda\omega^2_{\bf Q}/(4|\Delta(T)|^2)}.
\end{equation}
Here, $\omega_{\bf Q}$ denotes the frequency of the phonon mode responsible for driving the lattice-driven CDW instability, $\lambda$ is
the dimensionless electron-phonon coupling constant,
and $|\Delta|$ represents the CDW gap. At low temperatures, where
$4|\Delta|^2\gg\lambda\omega_{\bf Q}^2$,
the amplitudon frequency approaches
$\Omega_{\rm A}\simeq\sqrt{\lambda}\omega_{\bf Q}$, whereas as the temperature approaches the CDW
transition temperature and $|\Delta|\rightarrow 0$,
the amplitudon frequency becomes
$\Omega_{\rm A}\simeq 2|\Delta|\rightarrow 0$. As shown in Fig.~\ref{AMFIG}, the temperature dependence of the $M_1$ mode observed in our Raman measurements is inconsistent with the prediction of Eq.~(\ref{amplitudon}), which exhibits pronounced softening of the amplitudon energy primarily near 100~K. Fundamentally, the amplitudon represents oscillations in the magnitude of the order parameter around the minimum of the free-energy landscape~\cite{littlewood81gauge,Pekker2015}, and therefore exhibits pronounced softening primarily near 100~K as the order parameter vanishes.  
\end{itemize}

The absence of a clearly resolved amplitudon in Raman measurement may be attributed to several possible mechanisms. On one hand, the amplitude-mode response (estimated to occur at approximately $300~\mathrm{cm}^{-1}$ at low temperatures) may be masked by the strong Raman signals of the $A_g^5$ and $A_g^6$ phonon modes, both of which exhibit pronounced spectral features in the $250$–$300~\mathrm{cm}^{-1}$ range, making it difficult to identify as a distinct spectroscopic feature. On the other hand, the excitonic condensate in the present system only partially gaps the Fermi surface, as evidenced by the low-temperature tunneling conductance shown in Fig.~\ref{SFIG3}(c), which exhibits a nonideal U-shaped gap profile with pronounced particle-hole asymmetry. The remaining ungapped electronic states may provide additional decay channels for amplitude fluctuations, leading to strong damping and a broadened amplitude-mode spectral response at relatively high energies. A similar difficulty arises in cuprate superconductors, where low-energy nodal quasiparticles associated with $d$-wave pairing can strongly damp amplitude fluctuations~\cite{PhysRevB.87.054503}, making a sharp Higgs-mode resonance difficult to resolve in its corresponding measurements~\cite{Chu2020}.  Nevertheless, the existence and spectroscopic identification of a well-defined amplitudon in excitonic insulators remain open questions that require further theoretical and experimental investigation.

\section{Phenomenological model of Raman coupling}

Unlike conventional lattice-driven CDWs, the density-wave order considered here originates from the electronically driven excitonic condensate, whose collective phase excitations may couple to Raman scattering through the electronic polarization associated with the excitonic order. Raman spectroscopy has been shown to probe collective phase modes in electronically driven CDWs, as reported in charge-stripe-ordered cuprates~\cite{PhysRevLett.96.137003}. Here, we introduce a phenomenological model to describe the Raman coupling of the pinned phason in monolayer 1$T'$-MoTe$_2$. Specifically, the formation of the excitonic density-wave order breaks the underlying spatial symmetries and can induce an electronic polarization along the density-wave modulation direction. The corresponding electronic polarization is expressed as~\cite{kittel1963quantum}
\begin{equation}
\mathbf P=
\sum_{\mathbf{k},s}
\mathbf d_{\bf k+Q_{\rm CDW},k}^{cv}\langle
c_{\mathbf{k}+\mathbf Q_{\rm CDW},s}^{\dagger}
v_{\mathbf{k},s}\rangle
+\mathrm{c.c.},
\end{equation}
where $\mathbf d_{\bf k+Q_{\rm CDW},k}^{cv}$ is the generally complex vector-valued interband dipole matrix element. Using the excitonic correlation introduced above 
$\langle
c_{\mathbf{k}+\mathbf Q_{\rm CDW},s}^{\dagger}
v_{\mathbf{k},s}\rangle
=
F_{\mathbf{k},s}
e^{i(\theta_0+\delta\theta)}$, one has 
\begin{equation}
\mathbf P
=
2\mathrm{Re}\left[
\boldsymbol{\mathcal D}
e^{i(\theta_0+\delta\theta)}
\right],\qquad \text{where} \quad
\boldsymbol{\mathcal D}
=
\sum_{\mathbf{k},s}
\mathbf d_{\bf k+Q_{\rm CDW},k}^{cv}F_{\mathbf{k},s}. \label{eq:SM_pinned_polarization}
\end{equation}
When the symmetry of the density-wave state allows a longitudinal polarization while suppressing transverse components, the induced polarization is predominantly aligned with the density-wave modulation direction,
\begin{equation}
\mathbf P
\simeq
P_\parallel\hat{\mathbf Q}_{\rm CDW},\quad \text{with}\quad
P_\parallel
=
2|\mathcal D_\parallel|
\cos(\theta_0+\delta\theta+\phi).
\end{equation}
The effective phase $\phi$ originates from the complex interband dipole matrix elements and describes the phase offset of the electronic polarization relative to the excitonic order parameter.

\begin{figure}[h]
\centering
\includegraphics[width=0.5\linewidth]{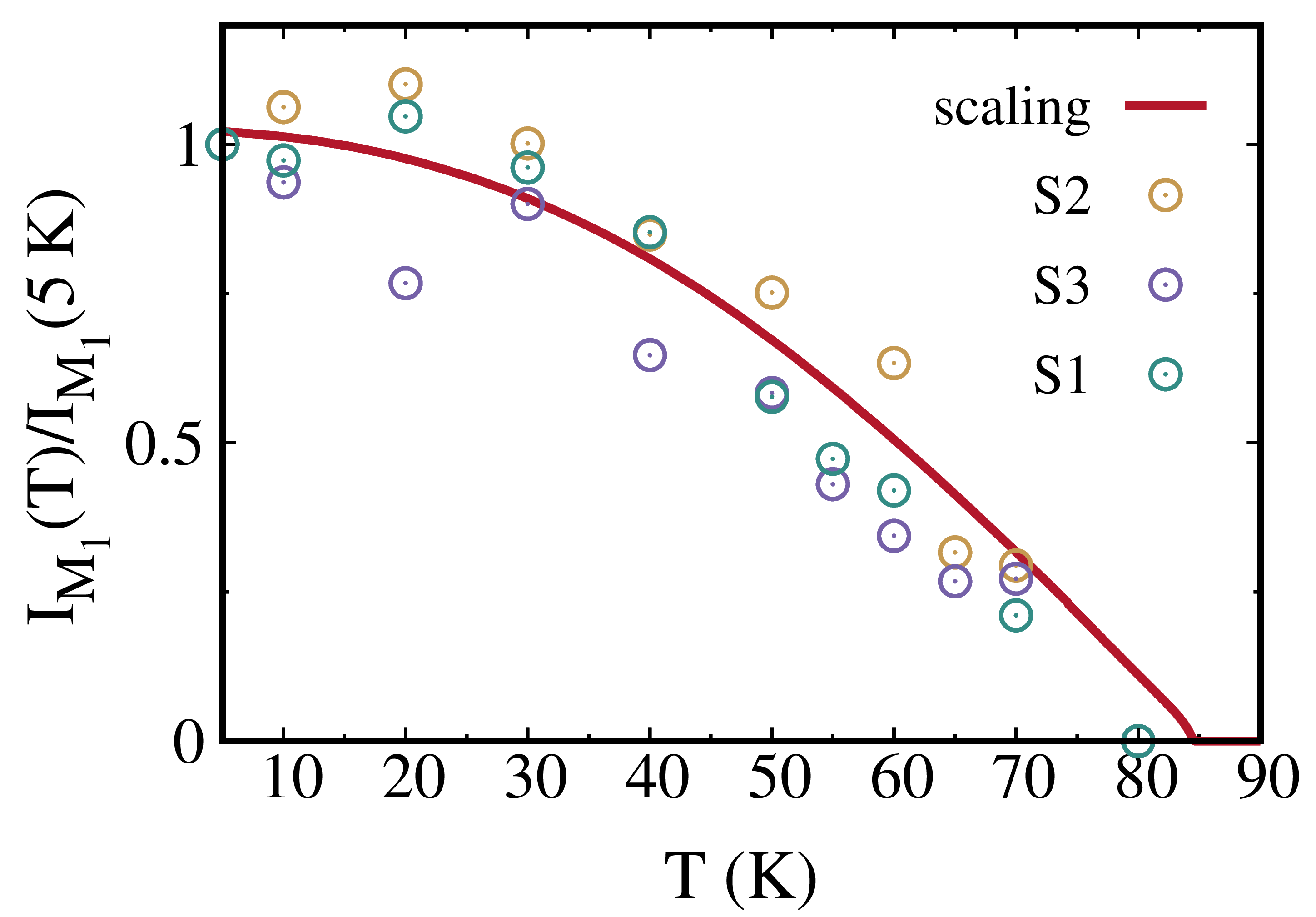}
\caption{Temperature dependence of the normalized  $M_1$ Raman intensity measured for three samples (S1, S2, and S3) in Fig.~1(f) of the main text, compared with the phenomenological scaling $I_{M_1}\propto\rho_{\mathrm{Q}}^4(T)$, corresponding to a Raman scattering amplitude $A_{M_1}\propto\rho_{\mathrm{Q}}^2(T)$. The value of $\rho_Q(T)$ is obtained from the self-consistent numerical calculation presented in Fig.~3(d) of the main text.}
\label{EMFIG}
\end{figure}

As derived in the preceding sections, thermal phase fluctuations suppress the observable density-wave order through the Gaussian renormalization factor $\exp[-\langle(\delta\theta)^2\rangle/2]$. Applying the same cumulant averaging to the electronic polarization gives
\begin{equation}
\overline{\mathbf P}
=
P_0(T)
\cos(\theta_0+\phi)
\hat{\mathbf Q}_{\rm CDW},
\end{equation}
where the coherent polarization amplitude is
\begin{equation}
P_0(T)
=
2|\mathcal D(T)|
\exp\Big[
-\frac12\langle(\delta\theta)^2\rangle
\Big]\propto|\Delta(T)|\exp\Big[
-\frac12\langle(\delta\theta)^2\rangle
\Big]\propto{\rho_{Q}(T)}.
\label{eq:SM_P0}
\end{equation}
To describe the corresponding Raman response, we consider the lowest-order symmetry-allowed quadratic contribution of the electronic polarization to the optical polarizability tensor, 
\begin{equation}
\alpha_{ij}
=
\alpha_{ij}^{(0)}
+
\beta_{ijkl}P_kP_l,
\label{eq:SM_Raman_polarizability}
\end{equation}
where $\beta_{ijkl}$ characterizes the coupling between the electronic polarization and the optical response. Substituting Eq.~(\ref{eq:SM_pinned_polarization}) gives
\begin{equation}
\alpha_{ij}
=
\alpha_{ij}^{(0)}
+
\beta_{ijkl}
\hat Q_k\hat Q_l
P_0^2(T)
\cos^2(\theta_0+\phi),
\end{equation}
where $\hat Q_i$ denotes the components of $\hat{\mathbf Q}_{\rm CDW}$. The Raman coupling to the collective phase fluctuation is determined by the derivative of the optical polarizability with respect to $\delta\theta$,
\begin{align}
\frac{\partial\alpha_{ij}}
{\partial\delta\theta}
&=
-\beta_{ijkl}
\hat Q_k\hat Q_l
P_0^2(T)
\sin\left[
2(\theta_0+\phi)
\right]. 
\end{align}
Then, the Raman scattering amplitude of the pinned phason for incident and scattered light with polarization vectors $\mathbf e^{\rm in}$ and $\mathbf e^{\rm out}$ is
\begin{align}
A_{M_1}(T)\propto
e_i^{\rm out}
\left.
\frac{\partial\alpha_{ij}}
{\partial\delta\theta}
\right|_{\rm eq}
e_j^{\rm in}\propto
e_i^{\rm out}
\beta_{ijkl}
\hat Q_k\hat Q_l
e_j^{\rm in}
P_0^2(T)\sin\left[
2(\theta_0+\phi)
\right].
\label{eq:SM_Raman_amplitude}
\end{align}
For a nonvanishing phase-sensitive Raman vertex, and assuming that the optical coupling coefficients and phase offset are approximately temperature independent, this directly yields 
\begin{equation}
A_{M_1}(T)\propto\rho_Q^2(T).
\end{equation}
Consequently, the Raman intensity approximately follows
\begin{equation}
I_{M_1}(T)
\propto
|A_{M_1}(T)|^2
\propto
\rho_Q^4(T).
\label{eq:SM_Raman_intensity}
\end{equation}
While the polarization-mediated Raman coupling introduced here is phenomenological, it provides a possible mechanism for the Raman activity of the pinned phason.  As shown in Fig.~\ref{EMFIG}, the resulting scaling $I_{M_1}\propto\rho_Q^4(T)$ captures the experimentally observed temperature dependence of the $M_1$ Raman intensity for the three samples, including its disappearance near the density-wave transition temperature. This provides a natural connection between the Raman response of the pinned phason and the long-range coherence of the excitonic density-wave order.

The polarization-mediated Raman coupling also provides a possible explanation for the $A_g$ symmetry of the $M_1$ mode. In particular, the excitonic density-wave order involves a modulation wave vector $\mathbf Q_{\rm CDW}$ predominantly along the $y$ direction, which breaks the corresponding spatial symmetries and allows an electronic polarization primarily along $y$. As a result, the polarization-mediated contribution to the optical polarizability in Eq.~(\ref{eq:SM_Raman_amplitude}) is dominated by the quadratic term $P_y^2$, i.e., 
\begin{equation}
A_{M_1}(T)\propto
e_i^{\rm out}
\beta_{ijyy}e_j^{\rm in}
P_0^2(T)\sin\left[
2(\theta_0+\phi)
\right].
\end{equation}
Under the $C_{2h}$ point-group symmetry, $P_y^2$ transforms as the fully symmetric $A_g$ representation. Consequently, the corresponding polarization-mediated contribution to the Raman response is associated with the $A_g$ symmetry channel, involving diagonal components of the optical polarizability, whereas the off-diagonal components characteristic of the $B_g$ representation are forbidden by the same symmetry constraints ($\beta_{xyyy}=0$). Such an $A_g$-type Raman response can be detected in both LL and LR circular-polarization configurations because of the anisotropy between the in-plane optical responses. This is consistent with the experimentally observed polarization dependence of the $M_1$ mode in Fig.~1(b). 

\section{Simulation details}
\label{sec_sd}

The electronic band structure and phonon spectrum of monolayer 1T$'$-MoTe$_2$ were calculated from first principles using the projector augmented-wave (PAW) method~\cite{blochl1994projector} implemented in the Vienna \textit{Ab initio} Simulation Package (VASP)~\cite{kresse1993abinitio}. The exchange-correlation interaction was treated using the Perdew-Burke-Ernzerhof (PBE) functional within the generalized gradient approximation (GGA)~\cite{perdew1996generalized}. Monolayer 1T$'$-MoTe$_2$ was modeled using a slab geometry with self-consistent spin-orbit coupling (SOC). Structural relaxation yields a lattice parameter of $b=3.468~\mathrm{\AA}$, corresponding to $\pi/b=0.9060~\mathrm{\AA}^{-1}$. Maximally localized Wannier functions were constructed from the SOC-included electronic structure using \textsc{Wannier90}~\cite{mostofi2008wannier90}, with initial projections onto Mo $d$ and Te $p$ orbitals.

\begin{figure}[h]
\centering
\includegraphics[width=0.5\linewidth]{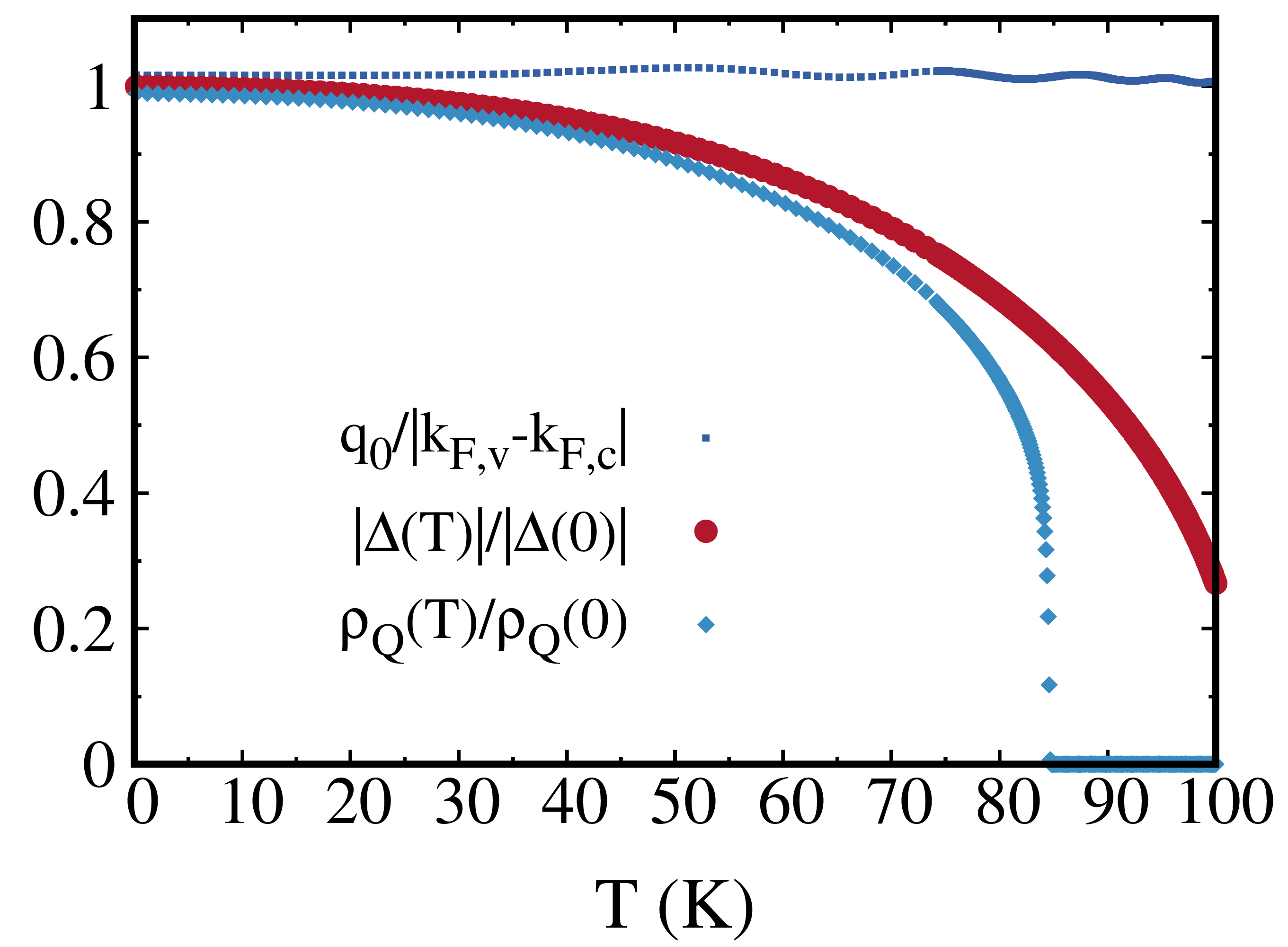}
\caption{Temperature evolution of the numerically calculated equilibrium pairing momentum $q_0$ and amplitude $|\Delta(T)|$ as well as $\rho_{\bf Q}(T)$.}
\label{AQFIG}
\end{figure}

For numerical calculations within our self-consistent microscopic theory, we first determine the pairing amplitude $|\Delta_{\mathbf q}|$ at each temperature by solving the gap equation [Eq.~(\ref{eq:SM_gap})]. The equilibrium pairing momentum $\mathbf q_0$ and amplitude $|\Delta(T)|$ are then obtained by maximizing $|\Delta_{\mathbf q}|$ with respect to $\mathbf q$ [Eq.~(\ref{eq:SM_q0})], which is equivalent to minimizing the mean-field free energy, as demonstrated in our previous theoretical study~\cite{yang2026excitonic}. Using the resulting $|\Delta(T)|$ and $\mathbf q_0$, we calculate the spatial phase stiffness $f^c_{ij}(T)$ [Eq.~(\ref{rhoij})] and the dynamical coefficient $D(T)=-\chi_{33}(T)$ [Eq.~(\ref{chi33def})], which determine the phason dispersion [Eq.~(\ref{phasonspectrum})]. The thermal phase fluctuations [Eq.~(\ref{eq:thph})] and the renormalized pinning gap $m_\theta(T)$ [Eq.~(\ref{phasonmass})] are subsequently determined self-consistently. Finally, the coherent density-wave amplitude [Eq.~(\ref{eq:SMDW})] is evaluated by incorporating the suppression of phase coherence due to thermal fluctuations. To model the electronic structure of the present material, we approximate the conduction and valence bands by isotropic parabolic dispersions,
\begin{equation}
\xi_{c,\mathbf{k}}=
\frac{\hbar^2k^2}{2m_c}-\frac{E_g}{2}-\mu,
\qquad
\xi_{v,\mathbf{k}}=
-\frac{\hbar^2k^2}{2m_v}+\frac{E_g}{2}-\mu.
\end{equation}
The band offset $E_g=210~\mathrm{meV}$ is taken from DFT calculations, while the effective masses $m_c=0.40m_e$ and $m_v=0.46m_e$ are extracted by fitting the DFT bands to isotropic parabolic dispersions, with $m_e$ denoting the free-electron mass. The chemical potential is set to $\mu=-40~\mathrm{meV}$. The effective interaction strength $g$ is determined by requiring the self-consistent solution of the gap equation to reproduce the experimentally estimated excitonic gap, $2|\Delta(T=10~\mathrm{K})|\approx38~\mathrm{meV}$ [Fig.~\ref{SFIG3}(c)]. The pinning strength is chosen such that $\sqrt{4l^2{\lambda}/D}=50~\mathrm{cm}^{-1}$. Furthermore, we restrict $\mathbf q\parallel\mathbf Q_e$, motivated by the DFT-calculated electronic band structure and Fermi-surface pockets [inset of Fig.~1(a) in the main text], which exhibit weak anisotropy that favors improved Fermi-surface nesting along this direction compared with the isotropic approximation~\cite{Qiu2026}. This anisotropy effectively selects the preferred direction of the pairing momentum~\cite{Qiu2026}, while its magnitude is primarily determined by the mismatch between the electron and hole Fermi surfaces. For the parameters considered, as shown in Fig.~\ref{AQFIG}, the self-consistent calculations yield $\mathbf q_0\approx(k_{F,v}-k_{F,c})\hat{\mathbf y}$ at all temperatures.

The DFT calculations yield an electronic valley momentum
$\mathbf Q_e=\pi/(3b)\hat{\mathbf y}$, where $b=3.468$~\AA.
Combining this with the self-consistently determined pairing
momentum $\mathbf q_0=0.04968\hat{\mathbf y}$~\AA$^{-1}$,
we obtain the density-wave ordering wave vector
\begin{equation}
\mathbf Q_{\mathrm{CDW}}
=\mathbf Q_e+\mathbf q_0
=0.35164\hat{\mathbf y}~\mathrm{\AA}^{-1},
\end{equation}
corresponding to a real-space modulation period of
$\lambda_{\mathrm{CDW}}=2\pi/|\mathbf Q_{\mathrm{CDW}}|
\approx5.16b$. This result agrees with the
 previously observed $5b$ periodicity~\cite{x6k9-wgk9,px77-3gg1}, corresponding to
$\mathbf Q_{\mathrm{exp}}=2\pi/(5b)\hat{\mathbf y}
=0.36235\hat{\mathbf y}$~\AA$^{-1}$,
with a relative deviation of approximately $2.96\%$.

\section{Methods for experimental measurements and data analysis}

\textit{Sample and device preparation.} MoTe$_2$ single crystals were grown using the self-flux method. Monolayer flakes for Raman spectroscopy and electrical transport measurements were mechanically exfoliated onto polydimethylsiloxane (PDMS) substrates inside a nitrogen-filled glove box. The layer number was confirmed by the presence of the CDW transition, which occurs exclusively in the monolayer limit~\cite{Qiu2026}. For Raman measurements, monolayer MoTe$_2$ was encapsulated with hexagonal boron nitride (hBN) and transferred onto a SiO$_2$/Si substrate. For transport measurements, the flake was picked up by hBN and assembled onto prepatterned graphite electrodes supported by hBN on SiO$_2$/Si. These structures were fabricated using a polycaprolactone (PCL)-assisted transfer process, with temperatures kept below 65~$^{\circ}$C. To investigate the influence of disorder on the CDW collective mode M$_1$, we employed thermal treatment, previously shown to increase disorder in MoTe$_2$~\cite{Wang2017}. Raman spectra were collected from the pristine sample (S1), followed by successive annealing at 150~$^{\circ}$C (S2) and 200~$^{\circ}$C (S3), each for 5~min in a nitrogen-filled glove box. The tunneling device was previously reported in Ref.~\cite{Qiu2026}. It was fabricated by sequentially picking up an approximately 10-nm-thick hBN encapsulation layer, a graphite tunneling electrode, a hBN tunneling barrier thinner than 2~nm, and a monolayer MoTe$_2$ layer, using a polycarbonate (PC)/PDMS stamp. The resulting heterostructure was deposited onto a clean, pre-fabricated bottom-gate structure on a Si/SiO$_2$ substrate to form the tunneling junction. Following assembly, the PC or PCL films were removed by rinsing with chloroform and isopropanol.

\textit{Raman scattering measurements.} Raman spectra were acquired in a backscattering configuration with 532-nm laser excitation focused through a 40$\times$ objective of a confocal microscope. The incident laser power was maintained below 0.05~mW to minimize sample heating. Scattered light passed through two Bragg notch filters before detection by a charge-coupled device coupled to a spectrometer. Unless otherwise specified for circular-polarization measurements, spectra were collected with parallel incident and scattered polarizations in the $c(bb)\bar{c}$ configuration to maximize the intensity of the $A_{\text{g}}^4$ phonon mode. A closed-cycle cryostat was used for temperature-dependent measurements.

\textit{Transport and differential conductance measurements.} Electrical resistance was recorded in a four-probe configuration using lock-in detection with a 1~$\mu$A excitation current at 17.77~Hz. For tunneling spectroscopy, a 0.1~mV ac voltage at 17.77~Hz was added to a variable dc bias. The differential conductance was determined from the resulting ac tunneling current measured by a lock-in amplifier.

\textit{Analysis of the Raman modes.} Raman peak parameters were determined by least-squares fitting after subtracting a linear baseline defined by the endpoints of each fitting window. Spectrally overlapping peaks, including M$_2$ and $A_{\text{g}}^4$, as well as $A_{\text{g}}^5$ and its higher-frequency neighbor, were fitted together. Other modes were fitted separately. The fitted parameters and associated uncertainties are reported in the main text and Supplemental Material.

\bibliography{ref.bib}